\documentclass[twocolumn,showpacs,preprintnumbers,amsmath,amssymb]{revtex4}
\usepackage [T2A]{fontenc}
\usepackage [cp1251]{inputenc}
\usepackage [english]{babel}
\usepackage [dvips]{graphicx}
\usepackage {array, tabularx, floatflt}

\begin{document}

\title{Energetics of charged metal clusters containing vacancies}

\author{Valentin V. Pogosov\footnote{Corresponding author:
vpogosov@zntu.edu.ua} and Vitalii I. Reva}

\address{Zaporozh'ye National Technical University, 69063 Zaporozh'ye, Zhukovsky Str. 64, Ukraine
\\
vpogosov@zntu.edu.ua}
\date{\today}

\begin{abstract}

We study theoretically large metal clusters containing vacancies. We
propose an approach, which combines the Kohn-Sham results  for
monovacancy in a bulk of metal and analytical expansions in small
parameters $c_{v}$ (relative concentration of vacancies) and
$R_{N,v}^{-1}$, $R_{N,v}$ being cluster radius. We obtain
expressions of the ionization potential and electron affinity in the
form of corrections  to electron work function, which require only
the characteristics of 3D defect-free metal.

The Kohn-Sham method is used to calculate the electron profiles,
ionization potential, electron affinity, electrical capacitance;
dissociation, cohesion and monovacancy-formation energies of the
small perfect clusters Na$_{N}$, Mg$_{N}$, Al$_{N}$  ($N \leq 270$)
and the clusters containing a monovacancy ($N\geq 12$) in the
stabilized-jellium model. The quantum-sized dependences for
monovacancy-formation energies are calculated for the Schottky
scenario and the ``bubble blowing'' scenario, and their asymptotic
behavior is also determined. It is shown that the asymptotical
behaviors of size dependences for these two mechanisms differ from
each other and weakly depend on the number of atoms in the cluster.
The contribution of monovacancy to energetics of charged clusters,
the size dependences of their characteristics and asymptotics is
discussed. It is shown that difference between the characteristics
for the neutral and charged cluster is entirely determined by size
dependences of ionization potential and electron affinity. Obtained
analytical dependences may be useful for the analysis of the results
of photoionization experiments and for the estimation of the size
dependences of the vacancy concentration including the vicinity of
the melting point.

\end{abstract}

\pacs {73.61.At, 36.40.Vz, 68.55.Ln, 68.65.Cd, 71.15.Mb, 73.30.+y,
32.10.Hq}

\maketitle

\section{Introduction}

Frenkel theory of melting assumes an abrupt increase of the
vacancies concentration at the triple point, as well as the decrease
in monovacancy-formation energies with increasing their
concentration \cite{179,Ubbelohde,Berry-2013}. The equilibrium
vacancy concentration is estimated from thermodynamic considerations
based on the monovacancy-formation energy, the magnitude of which,
in turn, can be extracted from positron annihilation spectroscopy
\cite{BPR-2015}. At the melting point, the relative concentration of
vacancies in metals is a fraction of a percent. Despite such small
concentrations, vacancies significantly influence properties of
solids. In the case of radiation damage in 3D metals or metal
islands, the vacancy concentration can be even tens of percent.

Small metal nanoclusters at low temperatures can be in
superconducting state, which results in a strong modification of the
energy spectrum. It is known that superconducting correlations
depend crucially on the density of states near the Fermi energy.
Certain shapes of nanoclusters support highly enhanced density of
states near electronic shell closings, see, e.g., recent works
\cite{KresinOvchin-2006,HalderKresin-2015}. For ideal nanoclusters,
the highest energy occupied electronic levels become strongly
degenerate at spherical shell closings (``magic''  numbers). The
presence of vacancies, as shown below, will lead to a change in the
magic numbers of atoms and might result in smearing of the effect of
shell structure on superconducting correlations including gap in
excitation spectrum.

Initially, in the experiments it was established that the
temperature melting of clusters on a substrate and free clusters
decreases with a decrease in their size \cite{Sambles,Borel}.
Interpretations of this mesoscopic phenomena were developed in a
number of papers
\cite{Yang-2007,Hendy,Guisbiers,Safaei,Luo,Chandra,qi-2016}. A
popular point of view is based on thermodynamical considerations:
near the temperature melting -- the smaller cluster size, the lower
a monovacancy-formation energy\, while the vacancy concentration is
not size- dependent \cite{Yang-2007,Guisbiers}.

Modern mass spectroscopic and calorimetric methods, allowing to
study in detail the process of premelting and postmelting in metal
clusters consisting of a countable number of atoms
\cite{Jarrold-2005,Haberland-2009,Jarrold-2010,Zamith}, have shown
that the melting temperature is characterized by an oscillatory size
dependence, and also has dimensional anomalies (for example, for
Al), poorly described by simple models
\cite{Jarrold-2005,Jarrold-2010}. Also, in the melting process the
diffusion of surface vacancies into bulk is more favorable for
clusters with unfilled electronic shells than for clusters with the
magic number of atoms \cite{Haberland-2009}. These facts stimulate
growing interest in the description of phase transition from the
solid to the liquid state, as the configuration excitation of
voids-vacancies in the clusters. For example, the question about the
size of the monovacancy-formation energies, the vacancy
concentration and a relation to the melting process remains open.

For the first time, mass-spectrometric measurements of dissociation
energy cluster ions Na$_{N}^{+}$ were reported  in
\cite{Brechignac}, and for Al$_{N}^{+}$ in \cite{Martin}.
Traditionally, according to such data, and also based on the
measured ionization potentials, the cohesion energy of neutral
clusters is calculated.

Recently, the ionization potential of Al$_{32-95} $ clusters and its
temperature dependence in the range 65 -- 230 K were measured in
\cite{Kresin-2015}. With increasing temperature, a decrease of the
ionization potential is not significant ($\sim 10 $ meV). The
melting point of such clusters is in the range 600 -- 700 K
\cite{Jarrold-2005,Jarrold-2010}.

The energy characteristics of solid clusters have been intensively
studied within different models and approaches, including triaxially
deformed ordinary jellium and \emph{ab initio} simulations (see
\cite{38,7,Landman-95,Vieira,Akola-2000,Kanhere,Aguado-2009} and
references therein). We use model of spherical stabilized jellium
\cite{25} that does not contain adjustable parameters. It is also
convenient and transparent for the analysis of the role played by
vacancies in cluster melting or other excitation processes.

However, with the exception of the works \cite{Alonso-88,Itoh},
self-consistent calculations for monovacancy-formation energy in
clusters and the impact on it of electronic spectrum quantization
has not yet been addressed. Therefore, one of the actual problems
that can be formulated in connection with melting of small-sized
aggregates, is the study of the size of its electron affinity and
ionization potential in the case of clusters contained vacancies. In
our model, the monovacancy is represented in the form of spherical
void of atomic size in a homogeneous positively charged background
due to the ions \cite{Alonso-88,Ziesche}.

An electron binding energy for small perfect clusters may be
calculated numerically only. In the opposite case of large clusters
the binding energy is close to the binding energy for the 3D metal
with the first size  correction due to surface curvature. In
\cite{Landman-95,Seidl-96,Seidl-1998} an exhaustive survey of the
semiclassical description of IP and EA in the form of the first
dimensional correction to the work function is given. In our work
the second size corrections are also taken into account and the
developed procedure is suitable for calculations of the ionization
potential and electron affinity of metal clusters containing point
defects and impurities. As an example, vacancies are considered.
This approach can be easily adapted for  calculation of the positron
attachment energy  for such clusters.

We propose a method, which combines several approaches. The first
one is based on a density-functional solution in the stabilized
jellium model for metal monovacancy in the bulk ignoring an external
surface. This problem was addressed by us earlier. We found a shift
in the energy of the ground state of the electrons \cite{BVP-2014-1}
and positrons \cite{BVP-2014-2} due to the presence of a
``subsystem'' of vacancies in 3D metal.

In the first approximation it is assumed that vacancies are
noninteracting because of their small concentration. However, there
are experiments showing that in some metallic systems the partial
ordering of vacancies is observed \cite{Amelinckx}, i.e. the energy
of vacancies formation has to be concentration dependent.

The second one uses a solution for a defect-free metal in presence
of an external flat surface, but with lower atomic density. The
lower density of the atoms is due to the existence of superlattice
of vacancies of relative concentration $c_{v}$  in the defected
metal. Using $c_{v}$ as a small parameter, all metal characteristics
are evaluated as series expansions. In these expansions the
zero-order terms are the characteristics of defect-free metal, and
the first-order terms are expressed through them. This approach
allows to obtain the vacancy dependence of electron work function
for 3D metal. These results are then used to evaluate an asymptotic
size dependence of the ionization potential of a spherical metal
cluster and the electron affinity.

The consistent procedure for the calculation of a size dependence of
ionization potential and electron affinity of a large spherical
metal cluster containing  vacancies is presented. In the framework
of effective medium approach, the perturbation theory over the small
parameters $c_{v}$, $R_{v}/R_{N,v}$ and $L_{v}/R_{v}$ is proposed
for the ionization potential and electron affinity ($R_{v}$ is the
average distance between vacancies and $L_{v}$ is the
electron-vacancy scattering length).

For small clusters Rb, K, Na, Li, Mg and Al containing monovacancy
we performed the Kohn-Sham calculations of the electron and
effective potential profiles; the ionization potential and an
electron affinity; the dissociation, the cohesion, the vacancy
formation energies. For demonstration and analysis of results, we
limited ourselves to metals Na, Mg and Al.

The size dependences for monovacancy-formation energies are
calculated for the Schottky scenario and the ``bubble blowing''
scenario, and their asymptotic behavior is also determined. It is
shown that the asymptotical behaviors of size dependences for these
two mechanisms differ from each other and weakly depend on the
number of atoms in the cluster. The contribution of monovacancy to
energetics of charged clusters, the size dependences of their
characteristics and asymptotics is discussed. It is shown that
difference between the characteristics for the neutral and charged
cluster is entirely determined by size dependences of ionization
potential and electron affinity.

\section{Ionization Potential and Electron Affinity}

\subsection{General relation}

By definition, the ionization potential and the electron affinity of
a metal cluster with the $N$ atoms  and radius
\begin{equation}
R_{N}  = N^{1/3}r_{0},\label{R1}
\end{equation}
where $r_{0}$ is the radius of Wigner-Seitz cell per unit atom, have
the form
\begin{equation}\label{1}
    \begin{aligned}
        & {\rm IP}_{N}  = E_{N}^{+}(R_{N})-E_{N}(R_{N}),\\
        & {\rm EA}_{N}  = E_{N}(R_{N})-E_{N}^{-}(R_{N}),
    \end{aligned}
\end{equation}
where $E_{N}^{+}\equiv E_{N}^{N_{e}-1}$, $E_{N}^{-}\equiv
E_{N}^{N_{e}+1}$ and  $E_{N}$ are total energies of charged and
neutral spheres; $N_{e}=ZN$ is the total number of electrons in a
neutral cluster, $Z$ is a valency of metal.

The basis of the semiclassical approximation is the expansion of the
electron chemical potential $\mu$ of valence electrons and the
surface energy per unit area $\sigma $ of the neutral cluster in
powers of the inverse radius $R_{N}^{-1}$ (liquid drop model)
\cite{1985,pppppp}:
\begin{equation}\label{Jell-51}
    \begin{aligned}
        & \mu(R_{N}) =  \mu _{0}+ \frac{\mu _{1}}{R_{N}}  + \frac{\mu
_{2}}{R_{N}^{2}}+O(R_{N}^{-3}),\\
        & \sigma(R_{N}) =  \sigma_{0}+ \frac{\sigma_{1}}{R_{N}}  +
\frac{\sigma_{2}}{R_{N}^{2}}+O(R_{N}^{-3}),
    \end{aligned}
\end{equation}
where $\mu _{0}$ and $\sigma_{0}$ correspond to flat surface
($R_{N}\rightarrow \infty $).

From the condition of mechanical equilibrium of the cluster, the sum
rules were obtained in Refs.
\cite{177,iakPog-1995,P-1995,keiPog-1996}. In particular
\begin{equation}\label{mu1}
    \begin{aligned}
        & \mu_{1}=\frac{2\sigma_{0}}{\bar{n}},\\
        & \mu_{2}=\mu_{1}\left(\delta_{1}-\frac{\sigma_{0}}{B_{0}}\right),
    \end{aligned}
\end{equation}
where $\delta_{1}=\sigma_{1}/\sigma_{0}$. Next, it is convenient to
use the values $\widetilde{\mu}_{1}\equiv \mu_{1}/r_{0}$,
$\widetilde{\mu}_{2}\equiv \mu_{2}/r_{0}^{2}$ having a dimension of
energy.

Consider a metallic cluster consisting of $N$ atoms and containing
$N_{v}$ vacancies. Then radius of the cluster is
\begin{equation}
R_{N,v}  = R_{N}\left(1+c_{v}\right)^{1/3}, \quad c_{v}=N_{v}/N.
 \label{111}
\end{equation}

Clusters of atoms have a structural periodicity, which is not
translational, but rather has a property of  ``spherical
periodicity''  \cite{Martin-1996}. This periodicity is due to
spherical layers of atoms (atomic shells or coordination spheres).
By this principle, one can also consider the vacancy ``subsystem''
in a cluster with distribution over spherical layers. We divide
conditionally the cluster into $N_{v}$ Wigner-Seitz supercells of
radius
\begin{equation}
R_{v}=\left(\frac{3}{4\pi n_{a}c_{v}}\right)^{1/3}\gg
r_{0}=\left(\frac{3}{4\pi n_{a}}\right)^{1/3}, \label{RR}
\end{equation}
where $n_{a}$ is the concentration of atoms. Respectively the
supercellular ``muffin-tin'' potential is replaced by a spherical
symmetric one.

For a large cluster with dilute subsystem of vacancies,  we will use
the expressions
\begin{equation}\label{Jell-52}
    \begin{aligned}
        & {\rm IP}_{N,v}  = W_{{\rm eff},v} - \frac{\mu _{1}}{R_{N,v}} -
\frac{\mu _{2}}{R_{N,v}^{2}}+
\frac{e^{2}}{2\mathcal{C}_{N,v}},\\
        & {\rm EA}_{N,v} = W_{{\rm eff},v}- \frac{\mu _{1}}{R_{N,v}} -
\frac{\mu _{2}}{R_{N,v}^{2}} - \frac{e^{2}}{2\mathcal{C}_{N,v}},
    \end{aligned}
\end{equation}
where $W_{{\rm eff},v}$ is the effective  electron work function of
3D metal containing vacancies, $-e$ is the electron charge. The last
term (\ref{Jell-52}) is the charging energy of  metallic sphere with
electric capacitance
\begin{equation}
\mathcal{C}_{N,v}= R_{N,v}. \label{CN}
\end{equation}

The analysis of experimental data for small metallic clusters is
usually carried out according to the formulae
\begin{equation}\label{Jell-alfa}
    \begin{aligned}
        & {\rm IP}_{N}  = W_{0}+  \frac{\alpha e^{2}}{N^{1/3}},\quad \alpha =\frac{1}{2r_{0}}-\frac{\widetilde{\mu}_{1}}{e^{2}},\\
        & {\rm EA}_{N}  = W_{0}-  \frac{\beta e^{2}}{N^{1/3}},\quad \beta
        =\frac{1}{2r_{0}}+\frac{\widetilde{\mu}_{1}}{e^{2}},
    \end{aligned}
\end{equation}
which will be used by us as reference expressions.

\subsection{Volume term for vacancy shift of the grand energy of electrons}

The translational symmetry of the lattice of a solid is at the basis
of calculation of the electron  work functions. Therefore, to take
into account the contribution of vacancies to work function, we have
to assume their periodic location in the form of a ``superlattice''
in a metal. In this case the vacancy shift of the bottom of the
electron conductivity band can be introduced.

The problem of description of our system within simple models is
that the bulk of the metal is not homogeneous due to vacancies, so
the density functional method in the jellium model must be solved as
a three-dimensional problem. Within the one-dimensional problem, it
is impossible to describe a set of spherically symmetric vacancies
and a planar external metal surface.

The wave function of the ground state of the electron is determined
by the Schr\"{o}dinger equation
\begin{equation}
  -\frac{\hbar^2}{2m}\nabla^{2}\Psi({\bf r}) + \Big[v_{\rm
eff}(r) +\sum\limits_{i=1}^{N_{v}}\delta v_{{\rm eff},v}(\textbf{r}
- \textbf{R}_{i})\Big]\Psi(\textbf{r}) = \varepsilon_{b} \Psi({\bf
r}), \label{k2}
\end{equation}
in which the vacancies are centered at the points $\textbf{R}_{i}$
($R_{i} <R_{N, v}$). One-electron effective potentials are
represented in such manner that the spherically symmetric potential
$v_{\rm eff} (r)$ forms the bottom of the conduction band and the
surface barrier, and $\delta v_{{\rm ef},v}(\textbf {r} - \textbf
{R}_{i}) $ forms the \emph{i}th vacancy (Fig. 1). In (\ref{k2}) the
energy $\varepsilon_{b}$ is a shift in the bottom of the conduction
band due to the presence of vacancies.

\begin{figure}[!t!b!p]
\centering
\includegraphics [bb = 27 294 567 794,width=0.48\textwidth] {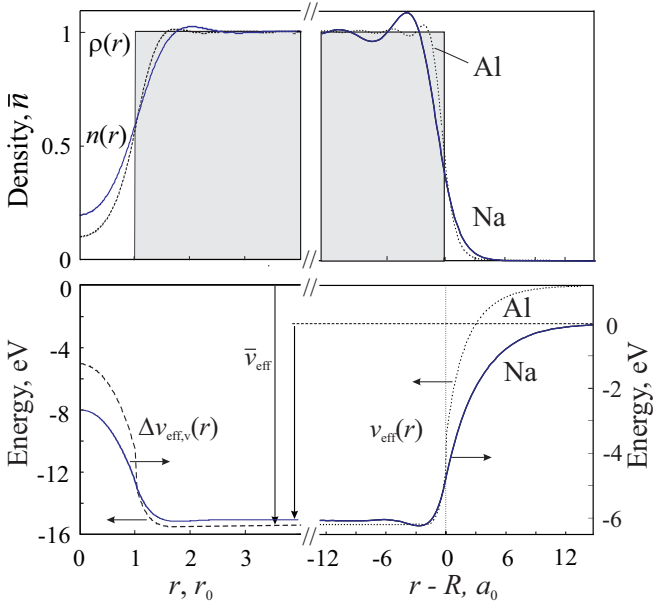}
\caption{The calculated in \cite{BVP-2014-1} electron and effective
potential profiles both near monovacancy  in the bulk and planar
surface ($R\rightarrow \infty$) for Na (blue solid line) and Al
(black dashed line).} \label{Fig1}
\end{figure}

Suppose that in 3D metal the potential field of the supercell has
translational periodicity. Then the electron wave function of the
ground state can be written in the form,
\begin{equation}
\Psi ({\bf r})=\psi({\bf r})u({\bf r}),  \label{k10}
\end{equation}
where the function $\psi({\bf r})$ (for 3D metal it is $e^{{\rm
i}\textbf{k}\textbf{r}}$) is modulated on the vacancy scale by the
function $u({\bf r})$. The function $u({\bf r})$ is the
superposition of the usual Wigner-Seitz solutions $u_{\rm WS}(|
\textbf{r}-\textbf{R}_{i}|)$ inside cells, centered at
$\textbf{R}_{i}$.

Next, $\varepsilon_{b}$ can be represented as the sum
\begin{equation}
\varepsilon_{b}=\varepsilon^{(0)}+\varepsilon^{(1)}, \label{Eb}
\end{equation}
\begin{equation}
\varepsilon^{(0)}=T_{0}+\langle \delta v_{{\rm
eff},v}\rangle_{R_{v}}, \label{tem3}
\end{equation}
where $T_{0}$ is the cellular energy eigenvalue; $\langle \delta
v_{{\rm eff},v}\rangle_{R_{v}}$ is averaged over the volume of the
supercell the contribution of potential energy from the
electron-vacancy potential $\delta v_{{\rm eff}, v} (r) = v_{{\rm
eff}, v}(r) - \bar{v}_{\rm eff}$, where $\bar{v}_{\rm eff}$ is the
position of the bottom of the conduction band in absence of
vacancies (Fig. 1).

The  value $T_{0}$ is determined from equation for supercell
\begin{equation}
\left[-\frac{\hbar ^{ 2}}{2m}\nabla^{2} +\delta v_{{\rm
eff},v}(r)-T_{0}\right] u_{\rm WS}(r)=0
  \label{k12}
\end{equation}
with the boundary conditions
\begin{equation}
\nabla u_{\rm WS}(r)|_{ r=R_{v}} =0,\,\,  u_{\rm WS}(r)|_{r=L_{v}}
=0,  \label{k13}
\end{equation}
where $L_{v}$ is the  scattering length of electron on the vacancy
(Table I). For an electron, the vacancy represents a potential
hillock, therefore $L_{v}> 0$.

\vspace*{.5cm}

%%%%%%%%%%%%%%%%%%%%%%%%%%%%%%%%%%%%%%%%%%%%%%%%%%%%%%%%%%%%%%%%%%%%
{\bf Table I.} The results of the Kohn-Sham calculation of $L_{v}$
for the electron scattering on monovacancy \cite{BVP-2014-1,Bab}.
%%%%%%%%%%%%%%%%%%%%%%%%%%%%%%%%%%%%%%%%%%%%%%%%%%%%%%%%%%%%%%%%%%%%%%%%%%%%%
\begin{center}
\begin{tabular}{ccccccccccc}     \hline\hline
Metal & Cs & Rb & K & Na & Li& Cu & Mg & Zn & Al & Pb \\
\hline
$Z$ &   1 & 1 & 1 & 1 & 1 & 2 & 2 & 3 & 3 & 4 \\

$r_{s},a_{0}$ & 5.63 & 5.2 & 4.86 & 3.99 & 3.28 & 2.11 & 2.65 &
2.31 & 2.07 & 2.30\\
$L_{v},a_{0}$ & 4.85 & 2.38 & 2.26 & 1.85 & 1.52 & 1.56 & 2.02 &
2.10 & 1.93 & 2.47\\
\hline\hline
    \end{tabular}
\end{center}
%%%%%%%%%%%%%%%%%%%%%%%%%%%%%%%%%%%%%%%%%%%%%%%%%%%%%%%%%%%%%%%%%%%%

\vspace*{.5cm}

Substitution of the expression for the electron wave function
\begin{equation}
u_{\rm WS}(r)=\frac{A}{\sqrt{4\pi }}\frac{\sin [q_{0}( r-L_{v})]%
}{q_{0}r}  \label{k15}
\end{equation}
in the second boundary condition gives the equation
\begin{equation}
\tan [q_{0}(R_{v} - L_{v} )] - q_{0}R_{v} = 0, \label{ViGz}
\end{equation}
from which we obtain
\begin{equation}
T_{0}=\frac{\hbar^{2} q_{0}^{2}}{2m}. \label{T00}
\end{equation}
This approach was first used in \cite{Spr}.

The interaction of an electron with vacancies can also be described
using the Fermi optical approximation
\begin{equation}
T_{0}=\frac{3\hbar^{2}L_{v}}{2mr_{0}^{3}}c_{v}. \label{k17}
\end{equation}

Here it should be noted that the smaller $c_{v}$ the better
agreement between results of calculations using (\ref{k15})
(\ref{T00}) and (\ref{k17}). Solving Eqs. (\ref{ViGz}) and
(\ref{k17}), for example, for Al with $c_{v} = 0.01$ leads to the
values $T_{0} = 0.0388$ eV and 0.0296 eV, respectively, which
evidences  that an accuracy of the optical approximation is lower.

The value of $T_0$ is determined by the  $s$-phase of scattering
($\rightarrow -L_{v} k$ as $k\rightarrow 0$) for $k = q_0$ and can
be tested by the value of residual resistivity of vacancies. All
phases of electron scattering for a monovacancy give contributions
to the resistivity, but the main contribution is produced by the
$s$-phase of scattering for $k = k_{\rm F}$ (see Table 1 in
\cite{BVP-2014-1}). The residual resistivity of vacancies estimated
in \cite{BVP-2014-1} turned out to be 2-3 times lower than the
experimental values.

The value  $\langle \delta v_{{\rm eff},v}\rangle_{R_{v}}$ in
(\ref{tem3}) is determined in the mean-field approximation by the
expression
\begin{equation}
\langle \delta v_{{\rm eff},v}\rangle_{R_{v}}=\frac{3c_{v}}{4\pi
r_{0}^{3}}\int_{0}^{R_{v}}dr\,4\pi r^{2}\delta v_{\rm eff,v}(r),
\label{DeltaWv}
\end{equation}
in which it is convenient to substitute  $R_{v}=r_{0}c_{v}^{-1/3}$.

Using the numerical solution of the problem of electron scattering
on the vacancy potential \cite{BVP-2014-1} we get
\begin{equation}
\varepsilon^{(0)}=A_{1}c_{v}+O(c_{v}^{2}), \label{full eff}
\end{equation}
where $A_{1}=4.10$ eV  and 13.3 eV for Na and Al, respectively.

We substitute (\ref{k10}) and (\ref{k12}) into the equation (\ref
{k2}), which after simple transformations can be rewritten in the
form
\begin{equation}
\left[-\frac{\hbar^{2}}{2m}\nabla^{2} -\frac{\hbar^{2}}{m}
\sum\limits_{i=1}^{N_{v}}\frac{\nabla u_{\rm WS}(\varrho)} {u_{\rm
WS}(\varrho)} \nabla + v_{\rm eff}(r)-\varepsilon^{(1)}\right]
\psi({\bf r}) =0, \label{k18}
\end{equation}
where $\varrho \equiv|\textbf{r}-\textbf{R}_{i}|$.

The  Eq. (\ref{k18}) contains the component of effective potential
in the form of a cross-term. It can be treated as a perturbation.
Earlier, analogous procedures were carried out in 3D perfect metals
for calculations of ground state energy $\varepsilon^{(0)}$ and
effective masses $m_{\rm eff}$ of electrons by Bardeen \cite{244},
Cohen and Ham \cite{245} and for positrons by Stott  and Kubica
\cite{246},
$$
\varepsilon^{(1)}=\frac{\hbar^{2}k^{2}}{2m_{\rm eff}}.
$$
In Refs. \cite{244} and \cite{245} the set of cellular wave
functions was used. In Ref. \cite{246} the function $\psi({\bf r})$
was expanded in the set of plane waves in a crystal.

It turned out that the influence of vacancies on effective masses of
electrons \cite{BVP-2014-1} and positrons \cite{BVP-2014-2} in
melting point ($c_{v}\approx 10^{-3}$) is insignificant. In order to
estimate the effective mass, we used in \cite{BVP-2014-1} the
Bardeen approach \cite{244} with $s$- and $p$-phase  of scattering
for $k=q_{0}$. For $c_{v}=10^{-2}$ the calculated values of the
electron effective mass $m_{\rm eff}=1.006$, 1.01, 1.012 and 1.014
for Na, Cu, Al and Pb, respectively. The small excess of $m_{\rm
eff}$ over $m$ indicates that the repulsion prevails over the
attraction in the scattering of electrons by the supercell potential
with the radius $R_{v}$.

\subsection{$c_{v}$-expansions}

We propose a method which combines two approaches. The first
approach is based on a self-consisten solution for the monovacancy
in 3D metal neglecting its surface. The second approach is based on
a stabilized jellium model for uniform 3D metal with flat boundary,
but the atom density of a metal is decreased due to the presence of
vacancies superlattice of relative concentration $c_{v}$. In this
case, the effective  electron work function $W_{{\rm eff},v}$ can be
presented as a sum
\begin{equation}
W_{{\rm eff},v}=W+\delta W_{v}^{\rm bulk}, \label{tem2}
\end{equation}
where $W$ is traditionally calculated by the density functional
method a characteristic consisting of a volume component and a
surface dipole barrier, and
\begin{equation}
\delta W_{v}^{\rm bulk}=-\varepsilon^{(0)}. \label{tem3333}
\end{equation}

For a semi-infinite metal [the $x$ axis is perpendicular to the
interface metal ($x\leq 0$) - vacuum ($x>0$)]  containing vacancies
the equilibrium profile $n(x)$ satisfies the the Euler-Lagrange
equation
\begin{equation}
\mu (x)= e\phi (x)+\left\langle \delta v\right\rangle _{\rm
WS}\theta (-x) + \frac{\delta G}{\delta n(x)}=\mbox{const},
\label{1.54}
\end{equation}
where the electrostatic potential $\phi(x)$ is determined by the
integration of the Poisson equation
\begin{equation}
\frac{d^{2}\phi}{dx^{2}}=-4\pi e[n(x)-\rho(x)], \quad
\rho(x)=\bar{n}\theta (-x). \label{8}
\end{equation}
Here $\mu$ is the electron chemical potential; $\left\langle \delta
v\right\rangle _{\rm WS}$ is the stabilization potential, $\theta
(-x)$ is the unit step function, $G[n]$ is the universal functional
corresponding to ordinary jellium model with energy per electron
$\varepsilon_{\rm J}=\varepsilon_{\rm t}+\varepsilon_{\rm xc}$  (for
detail see Ref. \cite{25}); distribution of the positive charge
$\rho ({\bf r}) $ is homogeneous inside the metal and zero outside
it.

The condition  $\mu(x)$ = const follows from the equivalence of the
choice of the coordinate in (\ref{1.54}). It is convenient to take
$x=-\infty$, where the gradient terms  (g. t.) vanish. Then the work
function of the electrons can be found as
\begin{equation}
W= -\bar{\mu}. \label{work}
\end{equation}

Let us make a comparison between a defect-free half-infinity metal
and a metal containing vacancies. The scenario for the formation of
vacancies is not important in this context. It may be the Schottky
mechanism or the ``bubble blowing'' mechanism \cite{Pog-94}. It is
important that the number of atoms in the sample is so large, that
in the spaces between vacancies a density of atoms is the same as in
perfect sample.

We average the charge density over the vacancy supercell. As a
result, the electroneutrality condition in the bulk of a
``fictitious, defect-free'' metal is
\begin{equation}
\bar{n} = \bar{ \rho}=\frac{Z\bar{n}_{a}}{1+c_{v}}, \label{volume}
\end{equation}
where  $Z$ is the valency of metal. The electron density satisfies
the condition $\frac{4}{3}\pi r_{s}^{3}\bar{n}=1$;
$r_{s}=r_{0}/Z^{1/3}$.

Carrying out the series expansion with respect to the small
dimensionless parameter $c_{v}$ in condition (\ref{volume}) and
limiting ourselves to linear terms, we have
\begin{equation}
\bar{n} = \bar{n}^{0}+\bar{n}^{1}c_{v},\quad
\bar{n}^{1}=-\bar{n}^{0}, \label{volume+}
\end{equation}
where $\bar{n}^{0}$ is the density of homogeneous electron gas for
$c_{v}=0$.

By analogy with the Refs. \cite{iakPog-1995,P-1995,keiPog-1996}, in
which a method is developed for determination of size corrections of
energy characteristics of spherical clusters in the stabilized
jellium ($R^{-1}$ is a small dimension parameter), we represent
characteristics of a metal containing vacancies as
\begin{equation}\label{expan}
    \begin{aligned}
        & n(x) = n^{0}(x)+n^{1}(x)c_{v}+\ldots,\\
        & \rho(x) = \rho^{0}(x)+\rho^{1}(x)c_{v}+\ldots,\\
        & \phi(x) = \phi^{0}(x)+\phi^{1}(x)c_{v}+\ldots,\\
        & \mu(x) = \mu^{0}(x)+\mu^{1}(x)c_{v}+\ldots.
    \end{aligned}
\end{equation}

This allows us to expand the equations (\ref{1.54}) and (\ref{8})
into series. Restricting ourselves to zero and first-order terms of
expansion (superscripts ``0'' and ``1'', respectively), we have
\begin{equation}
\mu^{0}(x)=e\phi^{0}(x)+\frac{\partial g^{0}}{\partial n^{0}}+
\left\langle \delta v\right\rangle_{\rm WS}^{0}\theta(-x)+\mbox{g.
t.}, \label{1.75-}
\end{equation}
\begin{equation}
\mu^{1}(x)=e\phi^{1}(x)+n^{1}(x)\frac{\partial ^{2}g^{0}}{\partial
(n^{0})^{2}} +\bar{n}^{1}\frac{\partial \left\langle \delta
v\right\rangle_{\rm WS}^{0}} {\partial \bar{n}^{0}}\theta
(-x)+\mbox{g. t.}, \label{1.75}
\end{equation}
\begin{equation}
\frac{d^{2}\phi^{0,1}(x)}{dx^{2}}=-4\pi e[n^{0,1}(x)-\rho^{0,1}(x)].
\label{8+}
\end{equation}
In (\ref{1.75}) $g^{0}\equiv n^{0}(x)\varepsilon_{\rm
J}\left(n^{0}(x)\right)$ is the energy density in LDA.

From condition  $x=-\infty$, we have sum-rules
\begin{equation}
\bar{\mu}^{0}=e\bar{\phi}^{0}(x)+\frac{\partial
\bar{g}^{0}}{\partial \bar{n}^{0}}+ \left\langle \delta
v\right\rangle_{\rm WS}^{0}=-W^{0}, \label{1.75--}
\end{equation}
\begin{multline}
\bar{\mu}^{1}=e\bar{\phi}^{1}+\bar{n}^{1}\frac{\partial
^{2}\bar{g}^{0}}{\partial (\bar{n}^{0})^{2}}
+\bar{n}^{1}\frac{\partial \left\langle \delta v\right\rangle_{\rm
WS}^{0}} {\partial \bar{n}^{0}} \\
=e\bar{\phi}^{1}
+\bar{n}^{1}\left[\bar{n}^{0} (\bar{\varepsilon}_{\rm SJ}^{0})''+
(\bar{\varepsilon}_{\rm J}^{0})'\right]=-W^{1}/c_{v}, \label{1.75+}
\end{multline}
where the energy per electron in the stabilized jellium
$\varepsilon_{\rm SJ}=\varepsilon_{\rm J}+\tilde{\varepsilon}$,
$\tilde{\varepsilon}$ is the electrostatic self-energy of the
uniform negative in the Wigner-Seitz cell \cite{25}.

Note that $\bar{\mu}_{0}$ in Eq. (\ref{Jell-51}) and $\bar{\mu}^{0}$
in Eq. (\ref{expan}) are the same, and  $\bar{\mu}_{1}$ and
$\bar{\mu}^{1}$ have different dimensions and are expressed in terms
of characteristics of a defect-free metal with a flat surface.

Integration in Eq. (\ref{8+}) gives
\begin{equation}
\bar{\phi}^{0,1} =-4\pi e \int\limits^{\infty}_{-\infty} dxx\left
[n^{0,1}(x)-\rho^{0,1}(x)\right]. \label{1.25a}
\end{equation}

Due to the fact that the vacancy shift of the electrostatic
potential in the bulk $\bar{\phi}^{1}$  is only due to a decrease in
average electron and ions densities in the metal, by analogy with
self-compressed clusters \cite{P-1995,keiPog-1996}, where the effect
is inversed, we have instead of (\ref{1.25a})
\begin{equation}
\bar{\phi}^{1} =\bar{n}^{1}(\bar{\phi}^{0})'. \label{1.25a+}
\end{equation}
Here the primes denote derivatives with respect to $\bar{n}^{0}$.

Using expression for the bulk modulus,
$B^{0}=(\bar{n}^{0})^{3}(\bar{\varepsilon}_{\rm SJ}^{0})''$, the
formulas (\ref{volume+}) and (\ref{1.25a+}), we finally have for
first-order term
\begin{equation}
W^{1}=-\left[e\bar{n}^{0}(\bar{\phi} ^{0})'+\left(B^{0}/\bar{n}^{0}+
\bar{n}^{0}(\bar{\varepsilon}_{\rm J}^{0})'\right)\right] c_{v}.
\label{1.72a}
\end{equation}

As a result of this approach
\begin{equation}
W_{{\rm eff},v}=W^{0}+W_{{\rm eff},v}^{1}, \label{tem2+}
\end{equation}
and vacancy contribution
\begin{equation}
W_{{\rm eff},v}^{1}=W^{1}+\delta W_{v}^{\rm bulk}, \label{1.72a-}
\end{equation}
as a whole is expressed only through the characteristics of a
defect-free metal. For defect-free Al and Na $W^{0}=4.30$ eV and
2.93 eV, respectively.

In the above exact formulas, the only $\bar{\phi}^{0}$ and
$(\bar{\phi}^{0})'$ terms require self-consistent calculations.
Table II shows the values of the components formula (\ref{1.72a}).
As we see, the contribution from the surface barrier $\sim
(\bar{\phi}^{0})'$ is very significant, competing with the bulk
contribution in (\ref{1.72a}) (a sum of terms in parentheses). In
general, $W^{1}$  are negative in sign. In almost all cases, the
major contribution of the effect is in the magnitude of $\delta
W_{v}^{\rm bulk}$ in (\ref{1.72a-}).

For Al at the melting point $c_{v}\approx 10^{-3}$. Consequently,
the contribution of equilibrium bulk vacancies to the electron work
function is approximately $-$0.2 eV. According to the paper
\cite{Gul},  the concentration $c_{v}$ should be much higher, then,
respectively, the effect of vacancies will increase by many times.

\vspace*{.5cm}

%%%%%%%%%%%%%%%%%%%%%%%%%%%%%%%%%%%%%%%%%%%%%%
{\bf Table II.} The results (in eV)  of the Kohn-Sham calculations
for componets of formula (\ref{1.72a}) and vacancy contribution
$W_{\rm eff}^{1}$ (\ref{1.72a-}) to the electron work function.
%%%%%%%%%%%%%%%%%%%%%%%%%%%%%%%%%%%%%%%%%%%%%%
\begin{center}
\begin{tabular}{cccccccccc}     \hline\hline
 Metal &  $e\bar{n}^{0}(\bar{\phi}
^{0})'$ & $B^{0}/\bar{n}^{0}$ & $\bar{n}^{0}(\bar{\varepsilon}_{\rm
J}^{0})'$ &$W^{1}/c_{v}$ & $W_{{\rm eff},v}^{1}/c_{v}$  \\
\hline
Na &  $-$0.773 & 1.77 & 0.0611 & $-$1.07 &$-$5.17 \\
Al &  $-$2.99 & 5.42 & 2.49 &   $-$4.92 & $-$18.2  \\
     \hline\hline
\end{tabular}
\end{center}
%%%%%%%%%%%%%%%%%%%%%%%%%%%%%%%%%%%%%%%%%%%%%%

\vspace*{.5cm}

Small independent parameters $R^{-1}$ and $c_{v}$ appear, when
applying these approaches to clusters.

\subsection{Size vacancy contribution}

Suppose that in large spherical clusters the potential field vacancy
suppercells has a ``spherical periodicity'' and function $\psi({\bf
r})$ in (\ref{k10}), which varies on the scale of the whole cluster,
is modulated by the function $u({\bf r})$. Then the boundary
conditions for $\psi({\bf r})$ on boundary  lead to a discrete
energy spectrum of electrons.

In the simplest case, assuming infinitely deep square well, the
energy of the ground state of electrons in the potential well
$v_{\rm eff}(r)$  of the whole cluster was found in \cite{PR-2017}.
This approach uses the boundary condition
\begin{equation}
\psi(r)|_{r=R_{N,v}}=0  \label{k21}
\end{equation}
for Eq. (\ref{k18}) and allows to derive an analytical expression
for vacancy quantum correction. The solution was obtained by
expanding the function $\psi({\bf r})$ in terms of the complete
orthonormal set of eigenfunctions corresponding to the deep
potential well. In this case, the term $\varepsilon^{(1)}$ in Eq.
(\ref{Eb}) can be written as
\begin{equation}
\varepsilon^{(1)}=\frac{\hbar ^{2}\pi ^{2}}{2mR_{N,v}^{
2}}+\left\langle \delta V(r)\right\rangle_{R_{N,v}}. \label{k22}
\end{equation}
The operator $\delta V(r)$ in (\ref{k18}) is used as a perturbation,
\begin{equation}
\delta V(r)=-\frac{\hbar ^{ 2}}{m}\sum\limits_{ i=1}^{
N_{v}}\frac{\nabla u_{\rm WS} (\varrho)}{u_{\rm WS}(\varrho)}\nabla.
\label{k23}
\end{equation}

The diagonal matrix element of the perturbation is the potential
field $\delta V(r)$,
\begin{equation}
\left\langle \delta V\right\rangle_{R_{N,v}}=\int \limits_{
r<R_{N,v}} d\textbf{r}\:\psi(r) \delta V(r)\psi(r), \label{k24}
\end{equation}
averaged over the ground state with the quantum numbers $n_{r}=1$,
$l=0$
\begin{equation}
\psi(r)=\frac{1}{\sqrt{4\pi }}\sqrt{\frac{2}{R_{N,v}}}\frac{\sin
(\pi r/R_{N,v})}{r}.
          \label{k25}
\end{equation}

As a result of the integration in Eq. (\ref{k24}) (see Appendix A),
using (\ref{mu1}), (\ref{Jell-52}), (\ref{k22})  and (\ref{k35}),
the final expression of the ionization potential becomes
\begin{multline}
{\rm IP}_{N,v}  = W_{{\rm eff},v}+\frac{\alpha
e^{2}}{N^{1/3}}\left(1-\frac{1}{3}c_{v}\right)
\\
+\frac{1}{N^{2/3}}\left[-\widetilde{\mu}_{2}\left(1-\frac{2}{3}c_{v}\right)\right.
\\
-\left.\frac{\hbar^{
2}\pi^{2}}{2mr_{0}^{2}}\left(1-\frac{2}{3}c_{v}-D_{1}\frac{L_{v}}{r_{0}}c_{v}^{1/3}\right)\right],
\label{Jell-IPv}
\end{multline}
where $D_{1}\approx 2.7$. Here we neglected the size dependence in
$\delta W_{v}^{\rm bulk}$ and the vacancy dependence in
$\widetilde{\mu}_{1}$ and $\widetilde{\mu}_{2}$. In our work
\cite{PR-2017} we used $\delta W_{v}^{\rm bulk}$ instead of $W_{{\rm
eff},v}$.

The corresponding expression for electron affinity ${\rm EA}_{N,v}$
is obtained from (\ref{Jell-IPv}) by replacing $\alpha\rightarrow
-\beta$.

In deriving $\varepsilon^{(1)}$, we used the boundary condition
(\ref{k21}). It does not take into account electron spillover. The
simplest way to account for the spillover is to use perturbation
theory \cite{Kubo-66}, taking into account the finite depth of the
spherical well   $\approx \bar{v}_{\rm eff}$, $\bar{v}_{\rm eff}=-6$
and $-16$ eV for Na and Al, respectively (Fig.1). There is no
analytical solution for a spherical well, and for a cluster-cube the
result is known \cite{PogKurbVas-2005}. This means that that result
$\sim \hbar^{2}$ in Eq. (\ref{Jell-IPv}) for hard wall-cube   must
be multiplied by
$$
\left(1- \frac{4}{R_{N}k_{0}}+O(R_{N}^{-2})\right),
$$
where  $k_{0}=\sqrt{-2m\bar{v}_{\rm eff}}$. The account of the
spillover will result in the appearance the term $-A/N^{1/3}$ in the
last parenthesis in the (\ref{Jell-IPv}), where $A\approx 1$ for Na
and Al. Thus it is obvious that in the derivation of the term we
neglected the terms proportional  $1/N$.

Therefore, taking into account that  $c_{v}\ll 1$, condition
(\ref{k21}) limits the value of $c_{v}$  from below. As a result of
the electron spillover  neglection, the applicability of formula
(\ref{Jell-IPv}) is possible under the condition
\begin{equation}
1/N\ll c_{v}=N_{v}/N\ll 1. \label{Kresin}
\end{equation}
The lower size limit where the concepts of supercells work is two
supercells shells ($N_{v}\geq 55$). The use of this technique for
one supercells shells ($N_{v}= 13$) is already an extrapolation.

The straightforward application of inequality (\ref{RR}) and the
boundary condition (\ref{k21}) in the paper \cite{PR-2017} leads to
the erroneous application of formula for ${\rm IP}_{N,v}$ to the
case of a monovacancy in a cluster (see Fig. 3 in \cite{PR-2017}).
The perturbation theory made it possible to introduce more realistic
description of the cluster. Because of this, without changing the
form of (\ref{k35}), the limits of applicability (\ref{Kresin}) of
the theory for a dilute subsystem of vacancies in a cluster are
formulated in the case of a discrete electron spectrum.

For $c_{v}=0.01$ and  the first two coordination spheres of
supercells, the total number of atoms in the cluster is $N= 1300$
and 5500,  which corresponds to $R_{N,v}\approx 1.8$ and 2.9 nm for
Al, respectively.

However, the choice of the applicability criteria of
(\ref{Jell-IPv}) is more consistent from the condition of expansion
of the electron chemical potential in power series of the inverse
radius ($c_{v}=0$)
\begin{equation}
\widetilde{\mu}_{1}N^{1/3}\geq \widetilde{\mu} _{2}+\frac{\hbar^{
2}\pi^{2}}{2mr_{0}^{2}}. \label{correct}
\end{equation}

In its turn, this leads to the values $N\geq 9.68\cdot 10^{3}$
$(R\approx 4.5$ nm)  and $N\geq 5.45\cdot 10^{4}$ ($R\approx 6$ nm)
for Na and Al, respectively.

The large clusters are detectable in the experiment. For example,
\cite{Hoffmann}  reported on a photoelectron spectroscopy and
determination of the steps of the Coulomb staircase
$(E_{C}=e^{2}/R_{N})$ of ionized clusters Al$_{N\leq 32000}^{-}$
with an accuracy of approximately $10^{-2}$ eV.

The value of  $\delta_{1}$ in (\ref{mu1}) was determined repeatedly
\cite{Pog-94}. We use calculated within a stabilized  jellium model
values $\delta_{1}=\sigma_{1}/\sigma_{0}=0.32r_{0},\,0.57r_{0}$ for
Na and Al, respectively [$\sigma_{1}=\gamma/2$ in Table VI of work
\cite{Ziesche}]. The experimental values of  $W_{0}=2.75$ eV,
IP$_{1}=5.14$ eV,  $\sigma_{0}/B_{0}=0.70$ $a_{0}$ for Na and
$W_{0}=4.28$ eV, IP$_{1}=5.99$ eV, $\sigma_{0}/B_{0}=0.40$ $a_{0}$
for Al are taken from \cite{86}.

In Fig. 2 the dependences (\ref{Jell-alfa}) and (\ref{Jell-IPv}) are
shown for Na and Al.  Due to the fact that the formula
(\ref{Jell-alfa}) gives the value IP$_{1}$  only 10\%  higher than
the experimental one, and the experimental values of $W_{0}$  for a
number of materials are strongly dependent on measurement
techniques, it was assumed from measurements IP$_{N\leq 100}$ for
magic numbers (see Fig. 28 in \cite{38}) using (\ref{Jell-alfa})  to
determine the true values $W_{0}$. On the other hand, the the last
term in (\ref{Jell-IPv}) that is conditioned by quantization lead to
non-linear behavior of IP$(N^{-1/3})$  and EA$(N^{-1/3})$  in the
region of small $N$.

\begin{figure}[!t!b!p]
\centering
\includegraphics [bb = 23 326 551 820,width=0.45\textwidth] {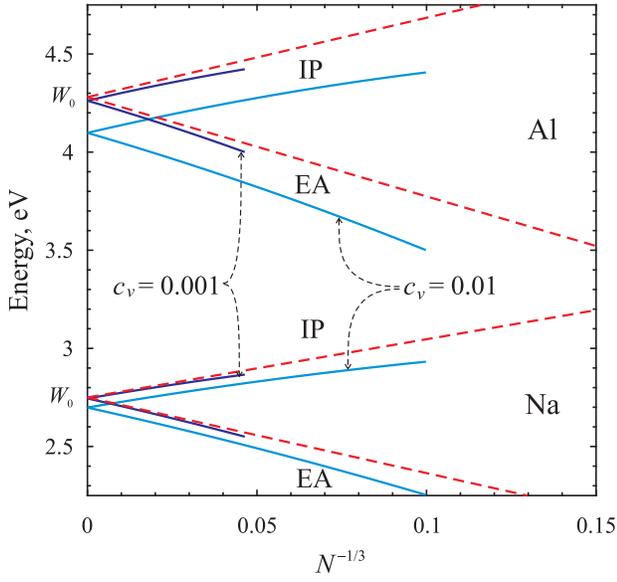}
\caption{The calculated dependences  IP$_{N,v}$ and EA$_{N,v}$ for
large clusters of Na and Al by the formulas: (\ref{Jell-IPv}) --
solid lines ($c_{v}=0.001$ and 0.01) under the condition
(\ref{Kresin}) and (\ref{Jell-alfa}) -- red dashed lines.}
\label{Fig2}
\end{figure}

The presented analytical approach seems to be promising for
experimental estimation of the concentrations of point defects or
impurities in metal clusters. To this end, it is first needed to
calculate the scattering length of electrons on the corresponding
defect in 3D metal. In particular, the problem of vacancies
concentration in the cluster at the melting point can be solved.
Using Fig. 2 as grid of values, if the experimental value of
IP$_{N,v}$ resides at one of the curves, the value $c_{v}(N)$  is
fixed for a given temperature.

\subsection{Small clusters containing monovacancy}

For simplicity, the density of the positive charge background of
defected and perfect (free-defect) clusters  is chosen in the form
\begin{equation}
\rho_{v}(r) = \bar{n}\theta(r-r_{0})\theta(R_{N,v}-r)  \label{rhho}
\end{equation}
and
\begin{equation}
\rho(r) = \bar{n}\theta(R_{N}-r), \label{rhhoho}
\end{equation}
respectively.

%%%%%%%%%%%%%%%%%%%%%%%%%%%%%%%%%%%%%%%%%%%%%%%%%%%%%%%%%%
\begin{figure}[!t!b!p]
\centering
\includegraphics [bb = 8 3 298 689, width=.48\textwidth] {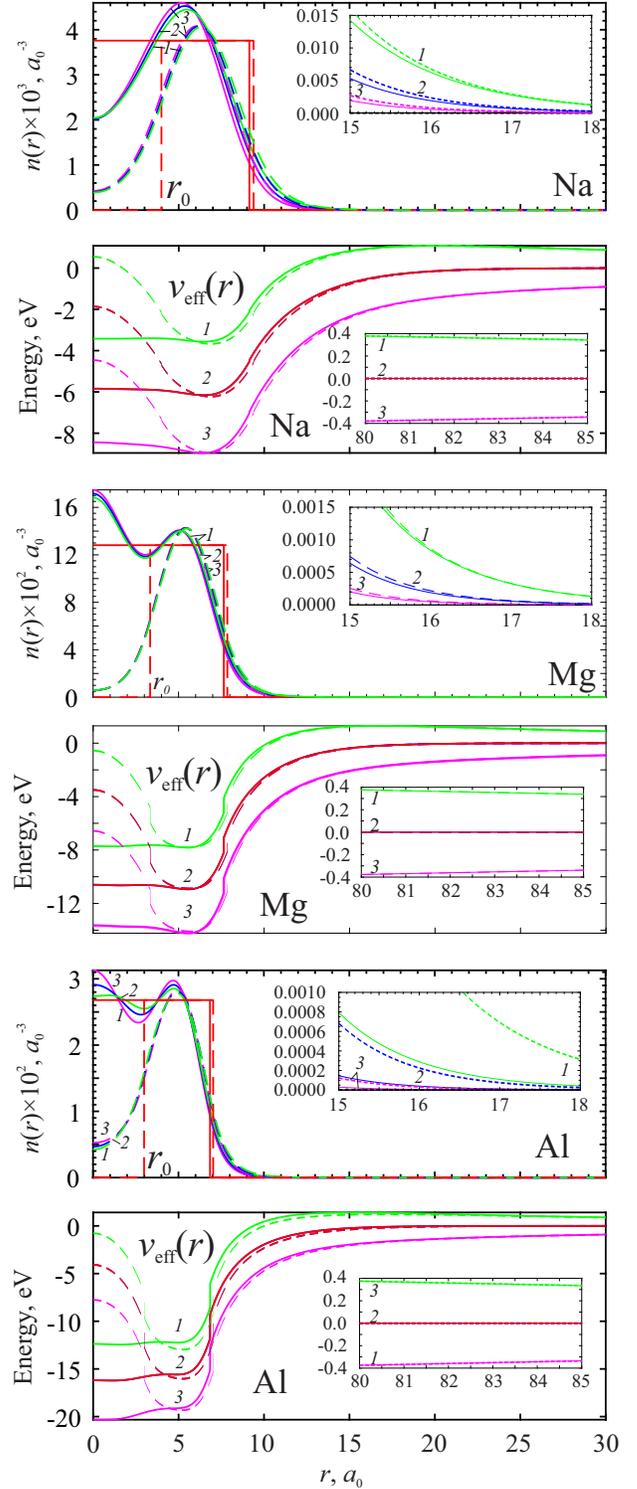}
\caption{Self-consistent profiles of spatial distribution of
electrons and effective potential for a charged and neutral perfect
clusters (solid lines), and a cluster with a monovacancy  (dashed
lines) containing the same number of atoms $N=12$; \emph{1} --
$Q=-e$, \emph{2} -- $Q=0$, \emph{3} -- $Q=+e$.} \label{Fig.3}
\end{figure}
%%%%%%%%%%%%%%%%%%%%%%%%%%%%%%%%%%%%%%%%

Wave functions and eigenvalues of energies $\varepsilon_{j,v}$ are
the solution of the Kohn-Sham equations with effective
single-electron potential, including electrostatic and
exchange-correlation potential in the LDA. The energy is measured
from the vacuum level, that is, on the energy of an electron with
zero kinetic energy, located far from the sample ($r \gg R_{N,v}$).

The electrostatic potential is a solution of the Poisson equation
for a fixed condition
\begin{equation}
\int_{0}^{\infty}dr\,4\pi r^{2}[\rho_{v}(r)-n_{v}(r)]=Q/e,
\label{neitr}
\end{equation}
where $Q$ is the total charge of the cluster.

%%%%%%%%%%%%%%%%%%%%%%%%%%%%%%%%%%%%%%%%%%%%%%%%%%%%%%%%%%
\begin{figure}[!t!b!p]
\centering
\includegraphics [bb = 4 4 286 537,width=.45\textwidth] {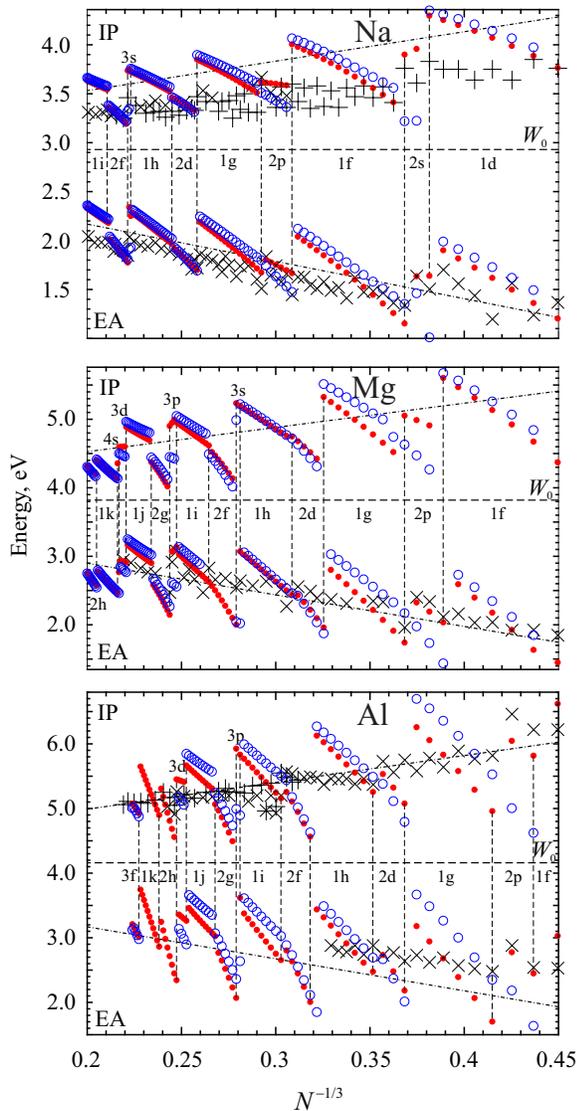}
\caption{The ionization potential IP and the electron affinity EA,
calculated from the formulas (\ref{1}) for perfect clusters (filled
red points) and clusters with monovacancy (open blue circles).
Experimental values: Na ($+$) \cite{38}, ($\times$) \cite{Kostko},
Mg ($\times$) \cite{Kostko}, and Al ($+$) \cite{Kresin-2015},
($\times$) \cite{38}. Dash-dotted lines are the asymptotics
(\ref{Jell-5+}). The letters s, p, d, f, g, h, i, j, k, l correspond
to the quantum orbital numbers $l=0,\ldots$, 9.} \label{Fig.4}
\end{figure}
%%%%%%%%%%%%%%%%%%%%%%%%%%%%%%%%%%%%%%%%

Fig. 3 shows the profiles of the electron distribution and effective
potential for a defect-free cluster and cluster with vacancy,
containing the same number of atoms $N = 12$. The vacancy radii for
Na and Al clusters are shown in the figure. Their radii differ
according to the definition (\ref{111}). The insets show the tails
of electron profiles and the potentials far from clusters. Despite
the fact that electronic distributions are rapidly decreasing
functions of  radial distance $r$, the potential tails extend far
(the calculation was carried out to  $r \approx R_{N}+900\,a_{0}$).
For charged clusters, the electrostatic potential decreases
asymptotically beyond $\sim Q/r$. It is interesting to compare the
profiles of electrons and potentials in Fig. 3 for Na$_{12}$ and
Al$_{12}$ with analogous profiles for 3D metals in Fig. 1. For large
clusters, the spatial profile becomes similar to the profile near
the surface of a semi-infinite metal, containing a large number of
the Friedel oscillations. Obtained profiles allow us to calculate
the total energy of the cluster $E^{\pm}_{N,v}$ (and $E^{\pm}_{N}$).

Fig. 4 shows the results of direct calculations of IP and EA by the
formula (\ref{1}), as well as the asymptotics (\ref{Jell-alfa}). A
difference between perfect clusters and defective clusters can be
understood from this figure as well as from Tables III and IV.  For
Na with the increase of $N$, starting from 12, this difference can
be 0.1 -- 0.5 eV (for Mg and Al about twice as much). Maximum
difference is observed at the transition from a completely filled
shell to an empty one. As $N$ increases, this difference is leveled.
For clarity, the results of the calculations are given in
coordinates $N^{-1/3}$. For clusters with a monovacancy $c_{v}=1/N$,
therefore there is a correspondence $N^{-1/3}=c_{v}^{1/3}$. In our
case $c_{v}\rightarrow 0$ for $N\rightarrow \infty$. If the vacancy
is not single, but their concentration is low (vacancies do not
interact with each other), according to our results, it is possible
to track the dependence of energy characteristics on the
concentration of vacancies.

The ionization potential and the electron affinity show a strong
oscillatory behavior due to the spherical shell structure. They tend
to $W_{0}$  asymptotically slow enough, due to high orbital
degeneracy and large angular quantum numbers $l$. According to the
results of experiments, the oscillations should be much weaker.
Going beyond LDA and using local spin density approximation (LSDA)
allows to reduce oscillations \cite{Ekardt-84}.

Using the Koopmans' theorem, the formulas (\ref{1})  can be
rewritten \cite{Perdew-88,Sabin} in form
\begin{equation}\label{Jell-5+}
    \begin{aligned}
        & {\rm IP}_{N,v}  = -\varepsilon_{N,v}^{\rm
HO}+\frac{e^{2}}{2\mathcal{C}_{N,v}^{+}},\\
        & {\rm EA}_{N,v}  = -\varepsilon_{N,v}^{\rm
LU}-\frac{e^{2}}{2\mathcal{C}_{N,v}^{-}},
    \end{aligned}
\end{equation}
where $\varepsilon_{N,v}^{\rm HO}$ / $\varepsilon_{N,v}^{\rm LU}$
and $\mathcal{C}_{N,v}^{\pm}$ are the energy of the upper occupied /
lower unoccupied orbital and electrical capacitances, respectively.

The spectra for perfect Na and Al clusters, shown in Fig. 5
demonstrate the filling of electronic shells as the number of
electrons is increased. For partially filled shells
$\varepsilon_{N,v}^{\rm HO}=\varepsilon_{N,v}^{\rm LU}\approx
\mu(R_{N,v})$. The maximum values of $\varepsilon_{N,v}^{\rm HO}$
correspond to completely filled shells, and magic numbers of atoms
$N^{*}$ for spherical perfect clusters and clusters with  vacancy
does not coincide in all cases. For Na the obtained values are
$N^{*}=$2, 8, 18, 20, 34, 40, 58, 68, 90, 92, 106, 132, 138, 168,
186, 196, 198, (230), 232, (252), 254; for Mg  $N^{*}=1$, 4, 9, 10,
17, 20, 29, 34, 45, 46, 53, 66, 69, 78, 93, 98, (99), \{115\},
(116), \{126\}, 127, 134, 153, 156, 169, 178, 199, 204, 219; for Al
 $N^{*}=$6, (30), 44, 46, 52, 62, \{66\}, (84), (102), \{104\},
\{136\}, 146, \{154\}, (180), (202), 204. In round brackets, the
obtained values are given for defective clusters which do not
coincide with the corresponding values for perfect clusters, and in
braces -- on the contrary.

As  $R_{N,v}$ increases, the values $\varepsilon_{N,v}^{\rm HO}$ and
$\varepsilon_{N,v}^{\rm LU}$, oscillate and tend to $\mu(R_{N,v})$
for $R\rightarrow\infty$. Amplitude of oscillations decreases
approximately as $R_{N,v}^{-3}$.

Let us denote the difference
$$ \Delta ({\rm IP}_{N})={\rm
IP}_{N,v}-{\rm IP}_{N}.
$$
and return to Fig. 4. At first sight, the sign $\Delta ({\rm
IP}_{N})>0$ is unexpected (circles are placed above the points for
the same $N$). Exceptions are clusters with such $N$ that maximum
contribution is given by levels with low $l$ ($ s, p$- and partially
$d$-orbital). In Fig. 4 these narrow areas are enclosed between
vertical dashed lines.

 \vspace*{.2cm}

%%%%%%%%%%%%%%%%%%%%%%%%%%%%%%%%%%%%%%%%%%%%%%%%%%%%%%%%%%%%%%%%%%%%
{\bf Table III.} Computed quantities ${\rm IP}$ and ${\rm EA}$ (in
eV) for Na clusters. Experimental values of ${\rm IP}_{\rm exp}$
(Na$_{17-94}$ -- \cite{38}, Na$_{132-140}$ -- \cite{Kostko}) and
${\rm EA}_{\rm exp}$ (Na$_{17-196}$ -- \cite{Kostko}).
%%%%%%%%%%%%%%%%%%%%%%%%%%%%%%%%%%%%%%%%%%%%%%%%%%%%%%%%%%%%%%%%%%%%
\begin{center}
\begin{tabular}{ccccccc}     \hline\hline
$N$ & ${\rm IP}_{N}$ & ${\rm IP}_{N,v}$ & ${\rm IP}_{\rm exp}$ &
${\rm EA}_{N}$ & ${\rm EA}_{N,v}$ & ${\rm EA}_{\rm exp}$ \\\hline
17 & 4.25 & 4.31 & 3.75 & 1.90 & 1.99 & 1.70 \\
18 & 4.30 & 4.35 & 3.83 & 1.64 & 1.01 & 1.50 \\
19 & 3.96 & 3.22 & 3.61 & 1.63 & 1.46 & 1.64 \\
20 & 3.90 & 3.22 & 3.76 & 1.15 & 1.35 & 1.34 \\
21 & 3.41 & 3.35 & 3.41 & 1.26 & 1.43 & 1.36 \\
22 & 3.49 & 3.46 & 3.57 & 1.35 & 1.40 & 1.46 \\
33 & 3.98 & 4.04 & 3.42 & 2.04 & 2.12 & 1.63 \\
34 & 4.00 & 4.06 & 3.60 & 1.66 & 1.47 & 1.45 \\
35 & 3.58 & 3.36 & 3.36 & 1.69 & 1.53 & 1.66 \\
36 & 3.59 & 3.40 & 3.57 & 1.72 & 1.58 & 1.58 \\
39 & 3.61 & 3.48 & 3.57 & 1.80 & 1.74 & 1.84 \\
46 & 3.64 & 3.71 & 3.48 & 1.89 & 1.98 & 1.63 \\
47 & 3.66 & 3.73 & 3.25 & 1.93 & 2.01 & 1.79 \\
48 & 3.68 & 3.75 & 3.43 & 1.96 & 2.04 & 1.75 \\
49 & 3.70 & 3.77 & 3.37 & 1.99 & 2.06 & 1.84 \\
57 & 3.83 & 3.88 & 3.44 & 2.19 & 2.24 & 1.86 \\
58 & 3.85 & 3.89 & 3.47 & 1.70 & 1.69 & 1.72 \\
59 & 3.33 & 3.31 & 3.33 & 1.72 & 1.72 & 1.81 \\
90 & 3.75 & 3.75 & 3.39 & 2.25 & 1.93 & 2.01 \\
91 & 3.75 & 3.34 & 3.40 & 2.34 & 1.91 & 2.03 \\
92 & 3.73 & 3.32 & 3.46 & 1.77 & 1.81 & 1.85 \\
93 & 3.19 & 3.22 & 3.29 & 1.79 & 1.83 & 1.87 \\
94 & 3.20 & 3.23 & 3.25 & 1.81 & 1.85 & 1.89 \\
131 & 3.66 & 3.69 & --- & 2.37 & 2.40 & --- \\
132 & 3.67 & 3.69 & 3.33 & 2.33 & 2.18 & 2.08 \\
133 & 3.62 & 3.46 & --- & 2.32 & 2.17 & --- \\
134 & 3.60 & 3.45 & --- & 2.31 & 2.17 & 2.18 \\
137 & 3.56 & 3.43 & 3.41 & 2.27 & 2.16 & 2.17 \\
138 & 3.54 & 3.43 & 3.46 & 1.91 & 1.95 & 1.96 \\
139 & 3.19 & 3.23 & --- & 1.92 & 1.96 & --- \\
140 & 3.20 & 3.23 & 3.21 & 1.93 & 1.97 & 1.99 \\
195 & 3.47 & 3.47 & --- & 2.26 & 2.25 & 2.14 \\
196 & 3.47 & 3.47 & --- & 2.17 & 2.15 & 2.16 \\
197 & 3.39 & 3.38 & --- & 1.99 & 2.16 & --- \\
252 & 3.72 & 3.72 & --- & 2.50 & 2.44 & --- \\
253 & 3.73 & 3.67 & --- & 2.50 & 2.42 & --- \\
254 & 3.73 & 3.64 & --- & 2.24 & 2.24 & --- \\
255 & 3.46 & 3.47 & --- & 2.24 & 2.25 & --- \\
256 & 3.47 & 3.47 & --- & 2.24 & 2.25 & --- \\
257 & 3.47 & 3.48 & --- & 2.24 & 2.25 & --- \\
258 & 3.48 & 3.48 & --- & 2.25 & 2.25 & --- \\
\hline\hline
    \end{tabular}
\end{center}
%%%%%%%%%%%%%%%%%%%%%%%%%%%%%%%%%%%%%%%%%%%%%%%%%%%%%%%%%%%%%%%%%%%%

From an analysis of the asymptotic behavior of ${\rm IP}_{N}$ and
${\rm IP}_{N,v}$ the basic vacancy dependence is contained in the
work function $W_{{\rm eff},v}<W_{0}$ (see (\ref{tem2+}) and Table
II).

In the case of small defected clusters, the perturbation of the
vacancy becomes essential, provided concentration $c_{v}\sim
R_{N}^{-3}$. It follows from Fig. 3 that the behavior of $v_{{\rm
eff},v}(r)$ is such that the electrons are squeezed out by a vacancy
from the center of the cluster to its surface and are grouped,
mainly, in the spherical layer $r_{0}<r<R_{N}$.  And when
integrating in spherical coordinates, the expression for the total
energy gives the main contribution to the energy. This is confirmed
by the spectral values of the energies, corresponding to the points
(circles) in Fig. 5. As an example, we give the values
$\varepsilon_{n_r,l}$ ($n_r$ and $l$ are the radial and orbital
quantum numbers)  for perfect and defective (in parentheses)
clusters: $\varepsilon_{0,0}=-4.925$ ($-$4.577),
$\varepsilon_{0,1}=-3.871$ ($-$3.831), $\varepsilon_{0,2}^{\rm
HO,LU}=-2.595$ ($-$2.708) eV for Na$_{12}$; and
$\varepsilon_{0,0}=-5.073$ ($-$4.755), $\varepsilon_{0,1}=-4.177$
($-$4.135), $\varepsilon_{0,2}^{\rm HO}=-3.119$ ($-$3.189),
$\varepsilon_{1,0}^{\rm LU}=-2.787$ ($-$2.048) eV for Na$_{18}$.

\vspace*{.5cm}

%%%%%%%%%%%%%%%%%%%%%%%%%%%%%%%%%%%%%%%%%%%%%%%%%%%%%%%%%%%%%%%%%%%%
{\bf Table IV.} Computed quantities ${\rm IP}$ and ${\rm EA}$ (in
eV) for Al clusters. Experimental values of ${\rm IP}_{\rm exp}$
(Al$_{20-63}$ -- \cite{38}, Al$_{89,90}$ -- \cite{Kresin-2015}) and
${\rm EA}_{\rm exp}$ (Al$_{20-23}$ -- \cite{38}).
%%%%%%%%%%%%%%%%%%%%%%%%%%%%%%%%%%%%%%%%%%%%%%%%%%%%%%%%%%%%%%%%%%%%
\begin{center}
\begin{tabular}{ccccccc}     \hline\hline
$N$ & ${\rm IP}_{N}$ & ${\rm IP}_{N,v}$ & ${\rm IP}_{\rm exp}$ &
${\rm EA}_{N}$ & ${\rm EA}_{N,v}$ & ${\rm EA}_{\rm exp}$ \\\hline
20 & 5.07 & 4.79 & 5.75 & 2.18 & 2.01 & 2.64\\
21 & 5.32 & 5.11 & 5.56 & 2.48 & 2.37 & 2.77\\
22 & 5.53 & 5.38 & 5.73 & 2.73 & 2.66 & 2.73\\
23 & 5.25 & 5.49 & 5.37 & 2.47 & 2.69 & 2.88\\
29 & 6.02 & 6.17 & 5.37 & 3.31 & 3.49 & ---\\
30 & 6.12 & 6.27 & 5.47 & 3.43 & 1.85 & ---\\
31 & 4.56 & 4.63 & 5.49 & 2.00 & 2.11 & ---\\
32 & 4.76 & 4.83 & 5.48 & 2.24 & 2.34 & ---\\
35 & 5.25 & 5.22 & 5.57 & 2.80 & 1.83 & ---\\
36 & 5.15 & 5.30 & 5.20 & 2.65 & 2.13 & ---\\
48 & 4.62 & 4.85 & 5.20 & 2.38 & 2.63 & ---\\
59 & 5.55 & 5.75 & 5.13 & 3.36 & 3.58 & ---\\
60 & 5.59 & 5.78 & 5.16 & 3.41 & 3.62 & ---\\
61 & 5.62 & 5.81 & 5.10 & 3.46 & 3.66 & ---\\
62 & 5.66 & 5.84 & 5.15 & 3.26 & 2.89 & ---\\
63 & 5.41 & 5.05 & 5.19 & 3.29 & 2.98 & ---\\
89 & 5.07 & 5.00 & 5.07 & 3.21 & 3.12 & ---\\
90 & 4.73 & 4.71 & 5.08 & 2.87 & 2.82 & ---\\
111 & 5.51 & 5.26 & --- & 3.75 & 3.52 & ---\\
112 & 5.53 & 5.24 & --- & 3.79 & 3.51 & ---\\
113 & 4.63 & 4.69 & --- & 2.91 & 2.99 & ---\\
114 & 4.67 & 4.74 & --- & 2.96 & 3.04 & ---\\
170 & 5.10 & 5.14 & --- & 3.75 & 3.81 & ---\\
171 & 5.13 & 5.16 & --- & 3.79 & 3.84 & ---\\
213 & 5.54 & 5.62 & --- & 4.56 & 4.67 & ---\\
214 & 5.57 & 5.65 & --- & 4.60 & 4.70 & ---\\
225 & 5.90 & 5.95 & --- & 5.00 & 5.11 & ---\\
226 & 5.81 & 5.81 & --- & 4.93 & 5.00 & ---\\
227 & 5.82 & 5.83 & --- & 4.97 & 5.04 & ---\\
230 & 5.86 & 5.86 & --- & 5.08 & 5.14 & ---\\
231 & 5.86 & 5.87 & --- & 5.11 & 5.17 & ---\\
263 & 6.44 & 6.36 & --- & 6.06 & 6.01 & ---\\
\hline\hline
    \end{tabular}
\end{center}
%%%%%%%%%%%%%%%%%%%%%%%%%%%%%%%%%%%%%%%%%%%%%%%%%%%%%%%%%%%%%%%%%%%%

%%%%%%%%%%%%%%%%%%%%%%%%%%%%%%%%%%%%%%%%%%%%%%%%%%%%%%%%%%
\begin{figure}[!t!b!p]
\centering
\includegraphics [bb = 13 1 266 583, width=.42\textwidth] {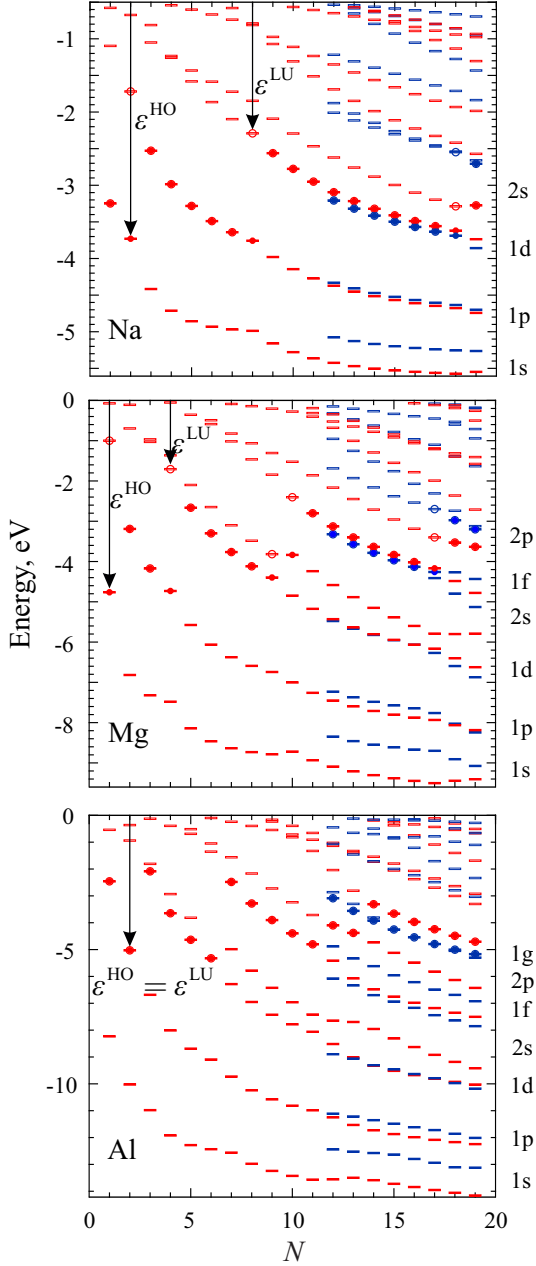}
\caption{Kohn-Sham one-electron eigenvalues for perfect (in red) and
defected (in blue) clusters. The  long solid and open rectangles
correspond to occupied and unoccupied levels, respectively. The
upper occupied $\varepsilon_{N}^{\rm HO}$ (filled red and blue
points) and lower unoccupied $\varepsilon_{N}^{\rm LU}$ (open red
and blue circles) energy levels are marked.} \label{Fig.5}
\end{figure}
%%%%%%%%%%%%%%%%%%%%%%%%%%%%%%%%%%%%%%%%

With increasing $N$, the contribution from the cluster bulk becomes
more and more important and for large enough $N$ the point and the
circles are interchanged, that is, the difference $\Delta ({\rm
IP}_{N\rightarrow \infty})$ becomes negative by the sign.

Self-consistent values of IP, EA, $\varepsilon^{\rm HO}$ and
$\varepsilon^{\rm LU}$, which are found according to the general
formulas (\ref{1}), allow us to evaluate capacitances of charged and
neutral clusters using the expressions
 (\ref{Jell-5+})
$$
\mathcal{C}_{N,v}^{+} = \frac{e^{2}}{2({\rm IP}_{N,v}
+\varepsilon^{\rm HO}_{N,v})},\quad \mathcal{C}_{N,v}^{-}  =
\frac{-e^{2}}{2({\rm EA}_{N,v} +\varepsilon^{\rm LU}_{N,v})},
$$
\begin{equation}
\mathcal{C}_{N,v}  = \frac{e^{2}}{{\rm IP}_{N,v} +\varepsilon^{\rm
HO}_{N,v}-{\rm EA}_{N,v} -\varepsilon^{\rm LU}_{N,v}}.
\label{capacity+-}
\end{equation}
Similar formulas for $\mathcal{C}_{N}$ correspond to a defect-free
clusters.

In classical electrostatics, the capacitance of conducting sphere is
determined by its radii $R_{N,v}$. Surface roughness on an atomic
scale (atoms have a finite volume) does not allow to determine
exactly the boundary \cite{Pog-94,P-1994}. In the jellium, the
boundary of the ion core always corresponds to the coordinate
$r=R_{N,v}$. However, the electronic cloud is more and more
``splashes out'' beyond the boundary of the core as its radius
$R_{N,v}$ decreases. Moreover, such a ``splashing'' depends on the
sign of the excess cluster charge (Fig. 3). As a result, the
quantities $\mathcal{C}_{N,v}^{+}$, $\mathcal{C}_{N,v}^{-}$ and
$\mathcal{C}_{N,v}$ are equal to each other only in the limit
$N\rightarrow \infty$.

%%%%%%%%%%%%%%%%%%%%%%%%%%%%%%%%%%%%%%%%%%%%%%%%%%%%%%%%%%
\begin{figure}[!t!b!p]
\centering
\includegraphics [bb = 57 24 514 836,width=.42\textwidth] {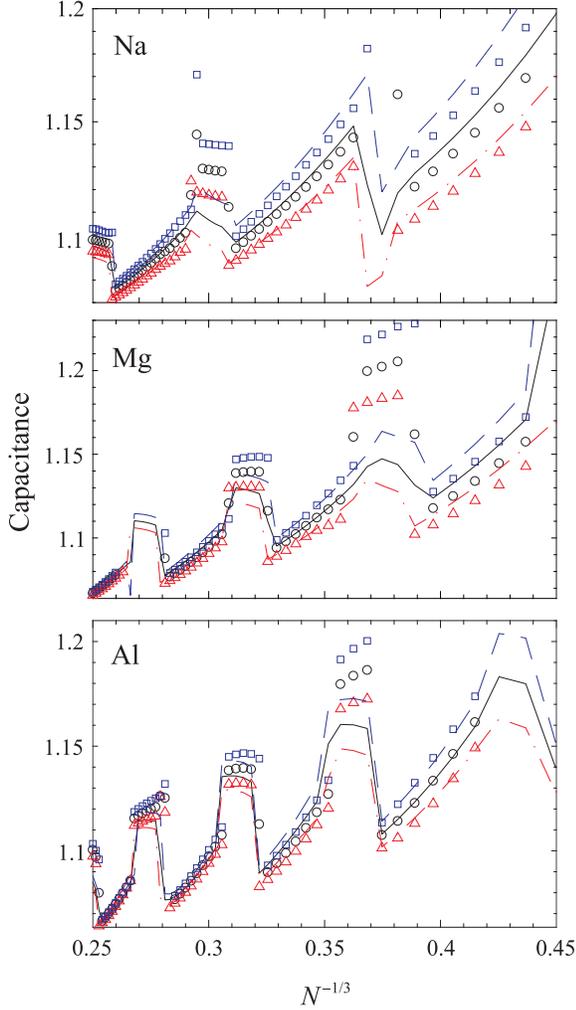}
\caption{The results of calculations by formulas (\ref{capacity+-})
for normalized capacitances of the perfect clusters (red
dash-dotted, black solid and blue dashed lines correspond to
positively charged, neutral, and negatively charged clusters,
respectively) and clusters with a monovacancy (red triangles, black
open circles, and blue open square, respectively).} \label{Fig.6}
\end{figure}
%%%%%%%%%%%%%%%%%%%%%%%%%%%%%%%%%%%%%%%%

Fig. 6 shows the results of calculations for capacitances
$\mathcal{C}_{N}$ and $\mathcal{C}_{N,v}$ normalized by their radius
$R_{N}$ and $R_{N,v}$  (atomic units), respectively.
Sign-alternating difference
$$
\Delta \tilde{\mathcal{C}}^{\pm}_{N}\equiv
\frac{\mathcal{C}_{N,v}^{\pm}}{R_{N,v}}-\frac{\mathcal{C}^{\pm}_{N}}{R_{N}}
$$
Is observed for certain intervals of $N$, which correspond to
filling of $s-$ and $p-$ electronic shells. The difference $\Delta
\tilde{\mathcal{C}}_{N}$ is determined mainly by the ratio of the
quantities $\varepsilon^{\rm HO}$ and $\varepsilon^{\rm LU}$ for
different $l$ in perfect and defective clusters, which can vary
depending on the radial quantum number. The capacitance of defective
clusters at filling shells with small $l$ is larger than for perfect
ones, and the opposite relation is observed for large $l$.

Using the experimental values of ${\rm IP}_{1}$ and ${\rm EA}_{1}$
for the Na atom ($R_{1}=r_{0}$), and also the condition
$\varepsilon^{\rm HO}_{1}=\varepsilon^{\rm LU}_{1}$ for unfilled
shells, as a test, we get the value $\mathcal{C}_{1}/r_{0}=1.8$. It
is in a satisfactory agreement with the calculated values for the
smallest clusters. For non-closed electronic shells, the cluster may
have a lower symmetry (spheroidal or triaxially deformed droplet
\cite{Landman-95}).

The charging effect expressed in the electric capacitance of the
cluster anions and cations depends on the sign of the excess charge.
Normalization of the capacitance  allows to give a simple
interpretation of the results of calculations: an excess negative
charge leads to an effective increase in the electron cloud
(effective radius) of the cluster, and excess positive charge
results in a decrease of both effective radius and capacitance. This
is qualitatively confirmed by the behavior of electronic profiles in
Fig. 3.

In \cite{Jarrold-2009}, in addition to measurements of thermal
capacity of cluster anions and cations Al$_{35-70}$, the ionization
potentials and electron affinity were calculated. Calculations are
carried out by the DFT under the condition of a global minimum of
the total energy for various configurations of atoms. The results
(in eV)  presented in Fig. 9 in \cite{Jarrold-2009} are approximated
by us in the form
\begin{equation}
 {\rm IP}_{N}=4.17+3.97/N^{1/3},
\label{Pasternak1}
\end{equation}
\begin{equation}
 {\rm EA}_{N}=3.88-3.05/N^{1/3}.
\label{Pasternak2}
\end{equation}
It follows from  (\ref{Jell-alfa}) that ${\rm IP}_{N}$ and ${\rm
EA}_{N}$ must approach the same quantity $W_{0}$ with increase of
$N$. Such a tendency does not agree with (\ref{Pasternak1}) and
(\ref{Pasternak2}). Also the comparison of (\ref{Pasternak1}) and
(\ref{Pasternak2}) from one side and (\ref{Jell-alfa}) from another
side allows to extract values of $\widetilde{\mu}_{1}$. The results
are ambiguous: it follows from (\ref{Pasternak1}) that
$\widetilde{\mu}_{1}\approx +0.58$  eV, while
$\widetilde{\mu}_{1}\approx -1.51$ eV from (\ref{Pasternak2}). Let
us remind that $\widetilde{\mu}_{1}\approx +0.68$ eV for Na
\cite{Seidl-1998,pppppp}.

\section{Dissociation and cohesion energies}

The dissociation energy of a neutral metallic (Me) cluster according
to the reaction Me$_{N}\rightarrow$Me$_{N-1}+$ Me$_{\rm at}$ is
determined by the difference in total energies
\begin{equation}
\varepsilon^{\rm dis}_{N}=[E_{N-1}+E_{\rm at}]-E_{N}=
N\varepsilon^{\rm coh}_{N}-(N-1)\varepsilon^{\rm coh}_{N-1}.
\label{Jell-2+}
\end{equation}
In the stabilized jellium, the energy of the atom $E_{\rm at}$ is
the total energy of a metal sphere of radius $r_{0}$.

By definition, the cohesive energy  $\varepsilon^{\rm coh}_{N}$ is
the binding energy (atoms in the cluster) per atom. It is determined
by the difference in the total energy of the $N$ free atoms and the
energy of a cluster consisting of $N$ atoms
\begin{equation}
\varepsilon^{\rm coh}_{N}=(NE_{\rm at}-E_{N})/N=E_{\rm at}-E_{N}/N.
\label{Jell-2}
\end{equation}
For $N\rightarrow \infty$, $\varepsilon^{\rm coh}_{N}\rightarrow
\varepsilon^{\rm coh}_{\infty}\equiv\varepsilon^{\rm coh}(r_{0})$.
The calculated values of $\varepsilon^{\rm coh}(r_{0})=$ 1.16, 1.17
 and 3.97 eV, respectively for Na, Mg and Al, are in satisfactory agreement
with the experimental values of $\varepsilon^{\rm coh}_{\infty}=$
1.11, 1.51 and 3.39 eV  (see Ref. \cite{Ziesche} and references
therein).

The binding equation has the form
\begin{equation}
\varepsilon^{\rm
coh}_{N}=\frac1N\sum\limits_{n=2}^{N}\varepsilon^{\rm dis}_{n}.
\label{Jell-20}
\end{equation}

Asymptotics of the size dependence of the cohesion energy
(\ref{Jell-2}) is well known \cite{179}
\begin{equation}
\varepsilon^{\rm coh}_{N}=\varepsilon^{\rm
coh}(r_{0})-\frac{2\sigma_{0}}{n_{\rm at}R_{N}}, \label{Jell-2 acc}
\end{equation}
where the last term can be written as  $-Z\mu_{1}/R_{N}$.

It should be noted that even in the works of Frenkel and Langmuir it
is noted that for some substances at low temperatures a universal
ratio is observed
$$
\frac{4\pi r_{0}^{2}\sigma}{q}  \approx \frac{2}{3},
$$
which is constructed from the observed values: the average distance
between atoms $r_{0}$, the surface energy $\sigma$, and the heat
evaporation $q=\varepsilon^{\rm coh}(r_{0})$ (see Table 2 in
\cite{Pog-94}). Using this relation, the asymptotics (\ref{Jell-2
acc})  can be rewritten in a form convenient for estimations
$$
\varepsilon^{\rm coh}_{N}\approx \varepsilon^{\rm
coh}(r_{0})\left(1-\frac{4}{9N^{1/3}}\right).
$$

Next, using Eqs. (\ref{Jell-2 acc}) and (\ref{Jell-2+}), we find a
coincidence of the asymptotics $\varepsilon^{\rm coh}_{N}$
(\ref{Jell-2 acc}) and $\varepsilon^{\rm dis}_{N}$.

It is of interest to determine the effect of charging on
dissociation and cohesion energies of clusters. The previous results
in electronic model and molecular dynamics simulations show that
fragmentation consists mainly in emission of single atoms
\cite{Bennemann}. Due to the fact that the ion work function is
greater than the heat of evaporation of the neutral atom
\cite{PogKurbVas-2005}, using Eqs. (\ref{Jell-2+}), (\ref{Jell-2})
and expressions
\begin{equation}
    \begin{aligned}
        & \varepsilon^{\rm dis,\pm}_{N}= N\varepsilon^{\rm coh,\pm}_{N}-(N-1)\varepsilon^{\rm coh,\pm}_{N-1},\\
        & \varepsilon^{\rm coh,\pm}_{N}=E_{\rm at}-E_{N}^{\pm}/N,
    \end{aligned}
\label{coh+-}
\end{equation}
the energy differences can be expressed as
\begin{equation}
 \begin{aligned}
 & \Delta\varepsilon^{\rm dis,+}_{N} \equiv \varepsilon^{\rm dis,+}_{N}- \varepsilon^{\rm dis}_{N}= {\rm IP}_{N-1}- {\rm IP}_{N},\\
 & \Delta\varepsilon^{\rm dis,-}_{N} =  {\rm EA}_{N} - {\rm
 EA}_{N-1},\\
        & \Delta\varepsilon^{\rm coh,+}_{N}=-\frac{1}{N}{\rm
        IP}_{N},\quad   \Delta\varepsilon^{\rm coh,-}_{N}=\frac{1}{N}{\rm EA}_{N}.
    \end{aligned}
\label{coh+}
\end{equation}

%%%%%%%%%%%%%%%%%%%%%%%%%%%%%%%%%%%%%%%%%%%%%%%%%%%%%%%%%%
\begin{figure}[!t!b!p]
\centering
\includegraphics [bb = 121 25 588 834,width=.45\textwidth] {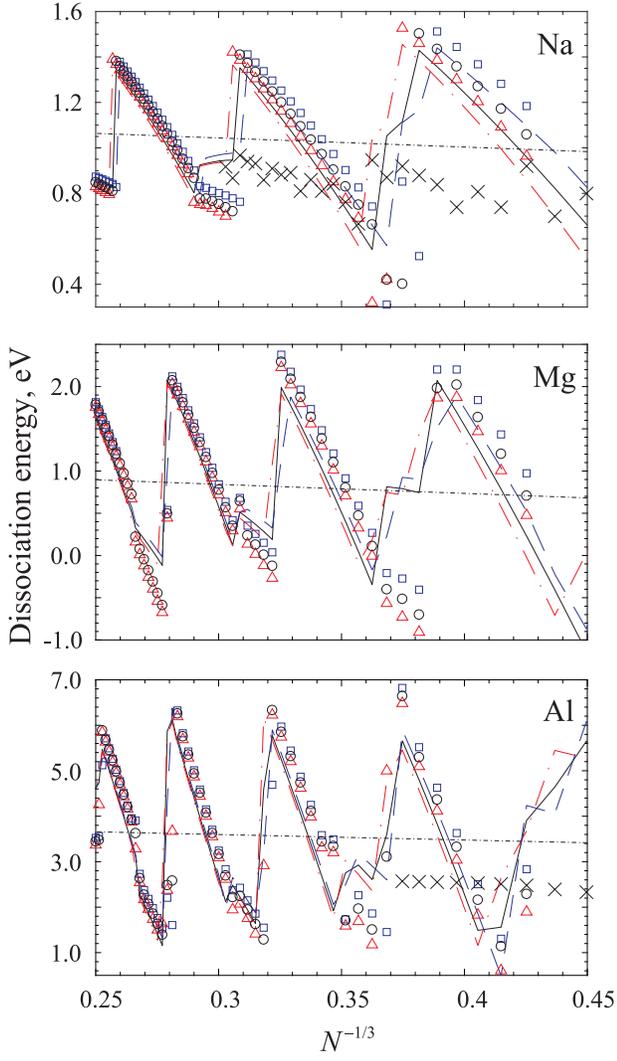}
\caption{The calculated dissociation energy $\varepsilon^{\rm dis}$
of the perfect clusters (red dash-dotted, black solid and blue
dashed lines correspond to positively charged, neutral, and
negatively charged clusters, respectively) and clusters with a
monovacancy (red triangles, black open circles, and blue open
square, respectively); experimental  ($\times$) values
\cite{Martin}; black dash-dotted lines are asymptotics (\ref{Jell-2
acc}).} \label{Fig.7}
\end{figure}
%%%%%%%%%%%%%%%%%%%%%%%%%%%%%%%%%%%%%%%%

%%%%%%%%%%%%%%%%%%%%%%%%%%%%%%%%%%%%%%%%%%%%%%%%%%%%%%%%%%
\begin{figure}[!t!b!p]
\centering
\includegraphics [bb = 14 11 442 833,width=.46\textwidth] {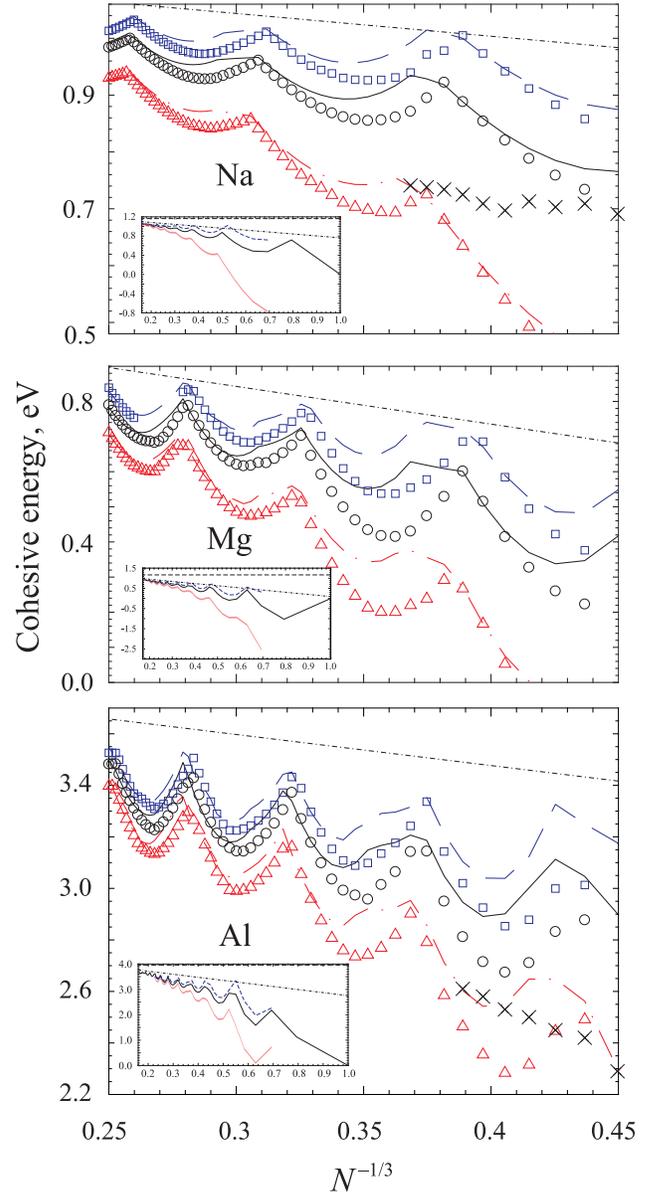}
\caption{Computed quantities $\varepsilon^{\rm coh}$ (red
dash-dotted, black solid and blue dashed lines correspond to
positively charged, neutral, and negatively charged clusters,
respectively) and clusters with a monovacancy (red triangles, black
open circles, and blue open square, respectively); experimental
($\times$) value \cite{Martin}; black dash-dotted lines --
asymptotics (\ref{Jell-2 acc}). The insets give the results for
perfect clusters in the whole range of $N$ under the study.}
\label{Fig.8}
\end{figure}
%%%%%%%%%%%%%%%%%%%%%%%%%%%%%%%%%%%%%%%%

Figs. 7 and 8 show the dissociation and cohesion energies of perfect
clusters and clusters containing monovacancy. Size dependence of the
dissociation energy in Fig. 7 is represented by quantum oscillations
around the asymptotical dependence. The values $\varepsilon^{\rm
dis}_{N,v}$ for defective cluster, with large $l$, are larger than
for a perfect one, and for small $l$, they interchange. For perfect
and defective clusters, in addition to a change of order filling of
electronic levels, a significant difference in the behavior of
dissociation energy with the same number of atoms $N$ is also seen:
for small $l$, dissociation energy of perfect clusters decreases
with increasing $N$, while it increases for defective clusters.
Comparison of the data in Fig. 8 and 9 confirms the accuracy of
formula (\ref{Jell-20}),  and also explains the difference in the
position of local maxima. The calculated values of $\varepsilon^{\rm
coh}_{N,v}$ are closer to the experimental values, obtained for
$T=150$ K \cite{Martin}, than $\varepsilon^{\rm coh}_{N}$. We note
that near the phase transition the quantities $\varepsilon^{\rm
dis}_{N}$, extracted from measurements of melting temperature and
latent heat of the transition can be negative \cite{Jarrold-2010}.

Thus, we can conclude that the most stable defect-free clusters are
those, which have latest filled levels with a small $l$, and for
defective one - on the contrary. In the experiments, the size
oscillations $\varepsilon^{\rm dis}_{N}$, are apparently suppressed
by temperature effects (see Fig. 9 in \cite{Brechignac}).

As shown in Fig. 8 cohesion energy of cluster anions and cations is
different from neutral clusters. Excess positive charge leads to a
decrease in the energy of cohesion due to the increase of the forces
of electrostatic repulsion, while excess negative charge leads to
the opposite effect. A behavior of dissociation and cohesion
energies of charged clusters is described by the Eq. (\ref{coh+}).

\section{Monovacancy-formation energy}

A considerable number of papers deal with first principles (\emph{ab
initio}) calculations of the monovacancy-formation energies in
metals \cite{Freysoldt-2014}. In the stabilized jellium and the
liquid drop models the energy of cohesion of an atom and the
monovacancy-formation energies are investigated in form of the
Pad\'{e} expansion \cite{Ziesche} (see also \cite{Pog-94}). Within
the notation of the work \cite{Pog-94}, the results of
\cite{Ziesche} can be expressed as
\begin{equation}
    \begin{aligned}
        & \varepsilon^{\rm coh}(r_{0})  = 4\pi r_{0}^{2} \sigma_{0}\left(1 +
          \tilde{\delta}_{1}+\tilde{\delta}_{2}\right),\\
        & \varepsilon^{\rm vac}(r_{0}) = 4\pi r_{0}^{2} \sigma_{0}\left(1 -
          \tilde{\delta}_{1}+\tilde{\delta}_{2}\right),
    \end{aligned}
\label{Jell-6 coh,vac}
\end{equation}
where $\tilde{\delta}_{1}\equiv\delta_{1}/r_{0}$ and
$\tilde{\delta}_{2}\equiv\delta_{2}/r_{0}^{2}$.

Calculated in \cite{BVP-2014-2}  $\varepsilon_{\infty}^{\rm
vac}\equiv \varepsilon^{\rm vac}(r_{0})=$ 0.33, 0.72 and 1.00 eV
agree with the experimental values of 0.335, 0.84 and 0.73 eV for
Na, Mg and Al, respectively \cite{Kraftmakher}.

Using values $\varepsilon^{\rm coh}(r_{0})$ and $\varepsilon^{\rm
vac}(r_{0})$, and $\tilde{\delta}_{2}=-0.13$ (Na), $-0.015$ (Mg) and
+0.22 (Al) from \cite{Ziesche}, we find $\tilde{\delta}_{1}=0.32$
(Na), 0.54 (Mg) and 0.57 (Al). The values $\delta_{1}$ and
$\delta_{2}$ are necessary for us in order to construct the
asymptotics of the monovacancy-formation energies.

For clusters, self-consistent calculations of $\varepsilon^{\rm
vac}_{N,v}$ have not been performed due to the need for detailed
description of the vacancy formation process. Therefore, it is of
interest to elucidate, which of the two scenarios of the vacancy
formation is favorable.

Within the Schottky scenario, an atom is extracted from the surface
of a perfect sphere, and in the final state the vacancy is in the
center of the sphere. In this case, we have
\begin{equation}
\varepsilon_{N,v}^{\rm vac,Sh}=[E_{N-1,\,v}+E_{\rm at}]-E_{N} =
N\varepsilon^{\rm coh}_{N} - (N - 1)\varepsilon^{\rm coh}_{N-1,\,v},
\label{Jell-3}
\end{equation}
where $E_{N-1,\,v}$ is the total energy of sphere with vacancy (the
spherical layer between $r=r_{0}$ and $r=R_{N-1,\,v}$ contains $N-1$
atoms).

According to another scenario \cite{Pog-94}, the number of atoms
does not change, but the ``bubble'' of radius $r_{0}$ is blown in
the center of the system. In this case, we have
\begin{equation}
\varepsilon_{N,v}^{\rm vac,blow}=E_{N,v}-E_{N} =
N\left(\varepsilon^{\rm coh}_{N} - \varepsilon^{\rm coh}_{N,\,v}
\right). \label{Jell-1}
\end{equation}

The comparison of Eqs. (\ref{Jell-3})  and (\ref{Jell-1})
demonstrates the advantage of the second mechanism by relation
\begin{equation}
\varepsilon_{N,v}^{\rm vac,Sh}=\varepsilon_{N,v}^{\rm
vac,blow}+\varepsilon^{\rm dis}_{N,v}. \label{Jell-4}
\end{equation}

We construct the asymptotics of the monovacancy-formation energy.
Its size dependence is determined by the difference in the total
energies of the spheres calculated by the formulas (\ref{Jell-3})
and (\ref{Jell-1})  in the limit $N \rightarrow\infty $, and reduces
to the difference of total surface energies.

For the vacancy blowing mechanism, using (\ref{111}), we obtain
\begin{multline}
\varepsilon_{N,v}^{\rm vac,blow}=4\pi R_{N,v}^{2} \sigma_{0}\left(1+
\frac{\delta_{1}}{R_{N,v}}+\frac{\delta_{2}}{R_{N,v}^{2}}\right)
\\
+\varepsilon^{\rm
vac}(r_{0}) -4\pi R_{N}^{2} \sigma_{0}\left(1 +
\frac{\delta_{1}}{R_{N}}+\frac{\delta_{2}}{R_{N}^{2}}\right)
\\
=\varepsilon^{\rm vac}(r_{0})\left(1+\frac{2}{3N^{1/3}(1 -
\tilde{\delta}_{1}+\tilde{\delta}_{2})}\right). \label{EvRblow}
\end{multline}
or the Schottky mechanism, in accordance with (\ref{Jell-4}) and
$R_{N-1,v}=R_{N}$, the asymptotics are determined by the sum of the
expressions (\ref{EvRblow}) and (\ref{Jell-2 acc}).  Asymptotic
dependence $\varepsilon_{N,v}^{\rm vac,Sh}$ weakly depends on $N$,
and the dependence (\ref{EvRblow}) demonstrates a decrease in
vacancy formation energy with an increase of $N$, which agrees with
the conclusions of the paper \cite{Sinder,Delavari}, but it
contradicts the conclusions of Refs.
\cite{Yang-2007,Hendy,Guisbiers}.

The asymptotic size behavior of the vacancy formation energy
(\ref{EvRblow}) can be qualitatively compared with the results of
\cite{Sinder} in which the energies of the cluster with or without
vacancy have been calculated on the basis of the tight-binding
approximation. Representing the expression (\ref{EvRblow}) in the
form
$$
\varepsilon_{N,v}^{\rm vac,blow}/\varepsilon_{\infty}^{\rm
vac}=1+C/N^{1/3},
$$
we obtained that the values $C = 1.21, 1.50$ and 1.03 for Na, Mg and
Al, respectively, are qualitatively consistent to $C = 1.33$ and
1.46 for Cu and $\beta-$Ti from \cite{Sinder}.

Fig. 9  shows the results of calculations of the
monovacancy-formation energy by two mechanisms. These calculations
confirm the formula (\ref{Jell-4}), namely, the advantage of blowing
a vacancy. All dependences experience strong fluctuations. For some
$N$, especially for Al, the values $\varepsilon_{N,v}^{\rm
vac,blow}$ become negative in narrow ranges of $N$. Such areas are
shown in Fig. 4 and above, the paper already contains comments on
hierarchy of electronic states in such clusters.

%%%%%%%%%%%%%%%%%%%%%%%%%%%%%%%%%%%%%%%%%%%%%%%%%%%%%%%%%%%
\begin{widetext}
\begin{figure}[h]
\center
\includegraphics[bb = 22 391 1147 1493,width=.91\textwidth]{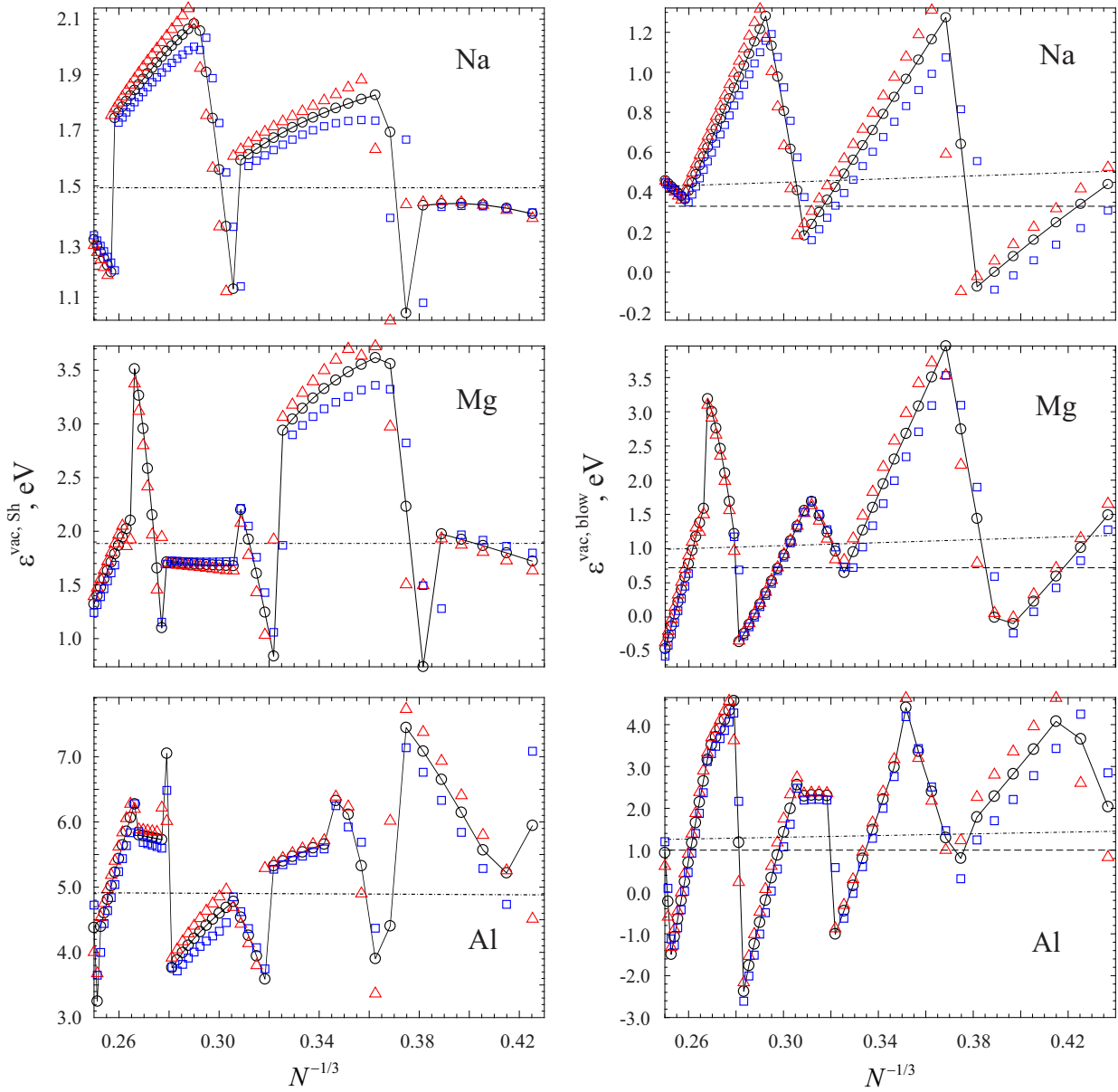}
\begin{minipage}{\textwidth}
\begin{flushleft}
\centering \caption{ Computed quantities $\varepsilon_{N,v}^{\rm
vac,Sh}$ (\ref{Jell-3}) and $\varepsilon_{N,v}^{\rm vac,blow}$
(\ref{Jell-1}) for clusters with a monovacancy (red triangles, black
open circles, and blue open square correspond to positively charged,
neutral, and negatively charged clusters, respectively); the black
dash-dotted lines are asymptotics (\ref{Jell-4}) and
(\ref{EvRblow}); horizontal black dashed lines (r. h. s.) correspond
to $\varepsilon_{\infty}^{\rm vac}$.}\label{Fig.9}
\end{flushleft}
\end{minipage}
\end{figure}
\end{widetext}
%%%%%%%%%%%%%%%%%%%%%%%%%%%%%%%%%%%%%%%%%%%%%%%%%%%%%%%%%%%%%%%%%%%%%

\vspace*{.2cm}

%%%%%%%%%%%%%%%%%%%%%%%%%%%%%%%%%%%%%%%%%%%%%%%%%%%%%%%%%%%%%%%%%%%%
{\bf Table V.} Monovacancy-formation energies (in eV) for neutral
Na$_{N}$.
\begin{center}
\begin{tabular}{|c|c|c|c|c|} \hline
N &  \multicolumn{2}{c|}{$\varepsilon_{N,v}^{\rm vac,Sh}$} &  \multicolumn{2}{c|}{$\varepsilon_{N,v}^{\rm vac,blow}$}\\
\cline{2-5} & This work  & \cite{Itoh} &  This work & \cite{Itoh}\\
\hline
55    & 1.80 &    1.35   &    0.49  & 0.49  \\
147   & 1.58  &    1.18   &   0.66   &  0.43/0.63  \\
\hline
\end{tabular}
\end{center}

%%%%%%%%%%%%%%%%%%%%%%%%%%%%%%%%%%%%%%%%%%%%%%%%%%%%%%%%%%%%%%%%%%%
\vspace*{.2cm}

The paper \cite{Itoh}  reports on the   results of ab initio
calculations on Na$_{N=55,147,309}$ clusters that show icosahedral
growth. The monovacancy-formation energy in cluster  depends on the
site at which the vacancy is created. A vacancy at the center or the
first atom shell is found to cost much higher energy compared to
other sites in the clusters.

In Table V, we compared the vacancy formation energy according to
the Shottky mechanism (the atom is removed from the cluster center
to outside) and according to the blowing bubble mechanism (the
removed atom is placed at the flat icosahedra surface \cite{Itoh}).
On one hand, the comparison of $\varepsilon_{N,v}^{\rm vac,Sh}$ (is
denoted by $E_{r}$ in Ref. \cite{Itoh}) reveals a difference of
about 0.4 -- 0.5 eV, but, on the other hand, it shows a similar size
dependence. Values of $\varepsilon_{N,v}^{\rm vac,blow}$ (is denoted
by $E_{v}$ in Ref. \cite{Itoh}) agree much better. In Table 5, in
order to demonstrate the dependence of $\varepsilon_{N,v}^{\rm
vac,blow}$ on the location of vacancy formation, two values are
presented, which are separated by the slash. They correspond to the
atom displacement from the cluster center and from the first
icosahedral atomic shell to the surface. The major aim of our simple
treatment is not on the absolute values of clusters characteristics,
but on their change upon vacancy formation.

The difference between the energies of formation of vacancies in a
charged and neutral cluster in accordance with two scenarios ($
E_{N}^{\pm}\rightarrow E^{\pm}_{N-1,\,v}+E_{\rm at}$ and $
E_{N}^{\pm}\rightarrow E^{\pm}_{N,\,v}$)  by analogy with
(\ref{coh+}) can be represented in the form of relations
\begin{equation}
    \begin{aligned}
        & \Delta\varepsilon_{N,v}^{\rm vac,Sh,+} =  {\rm IP}_{N-1,v} - {\rm IP}_{N},\\
        & \Delta\varepsilon_{N,v}^{\rm vac,Sh,-} = {\rm
        EA}_{N} - {\rm EA}_{N-1,v},\\
        & \Delta\varepsilon_{N,v}^{\rm vac,blow,+} =  {\rm IP}_{N,v} - {\rm IP}_{N},\\
        & \Delta\varepsilon_{N,v}^{\rm vac,blow,-} = {\rm
        EA}_{N} - {\rm EA}_{N,v}.
    \end{aligned}
\label{vac+}
\end{equation}
The character of the size dependence $\varepsilon_{N,v}^{\rm
vac,\pm} $ (\ref{vac+}) is fully supported by direct calculations
 and is determined by the behavior of IP and EA in
Fig. 4.

The above calculations in LDA correspond to zero temperature.
Perhaps, for a density of atoms corresponding to finite temperature,
reduced symmetry of clusters, as well as the use of LSDA for
exchange-correlation energy, strong oscillatory behavior of the
energy characteristics will be suppressed. Anyway, the size behavior
of the results of direct calculations agrees with its asymptotics.

In quasi-thermodynamics, the probability of the appearance of
vacancies in a cluster at finite temperature $T$, can be estimated
from the condition of free energy variation $\Delta F_{N,v}^{\rm
vac,blow}$,
\begin{equation}
\Delta F_{N,v}^{\rm vac,blow}=\varepsilon_{N,v}^{\rm
vac,blow}-T\Delta S_{N,v}^{\rm vac,blow}\leq 0. \label{Freeenergy}
\end{equation}

As a result of the fact that when the vacancy is blown, the number
of ions in the cluster does not change, the entropy contribution is
provided only by the degenerate electron gas \cite{A-78}. The
corresponding expression is
\begin{multline}
T\Delta S_{N,v}^{\rm vac,blow}= \frac{2\pi^{5/3}}{3^{2/3}}
\left(\frac{k_{\rm B}T}{e^{2}}\right)^{2}
\\
\times\int_{0}^{\infty} dr\,
r^{2}\left[n_{N,v}^{1/3}(r)-n_{N}^{1/3}(r)\right]. \label{DeltaS}
\end{multline}

For calculations in (\ref{DeltaS}), equilibrium profiles of electron
distributions in the stabilized jellium will be needed for given $N$
and $T$. At zero temperature and $N = 12$ these profiles are
presented on Fig. 3.

\section{Summary}

In this paper, we propose a consistent formal procedure for finding
ionization potential of a large metal cluster containing the
vacancies that is based on a previously solved problem
 on the scattering of electrons by a monovacancy in bulk metal
by the Kohn-Sham method in the stabilized jellium. Self-consistent
profiles are used to determine the vacancy energy shift of the
ground state in a metal cluster sphere as a series of size
corrections.  Spherical periodicity in the arrangement vacancies was
assumed. The limits of applicability of this expansion to powers of
the inverse radius are $ R> 4.5$ nm and $R> 6$ nm for Na and Al,
respectively.

In our work, the effect of bulk vacancies on the properties of a
cluster is estimated. Only by introducing periodic arrangement of
internal vacancies it is possible to consider the defected sample as
a cellular media and to estimate the shift of its ionization
potential. The presented approach seems to be promising for
experimental estimation of the concentrations of bulk point defects
or impurities in metal clusters. To this end, it is first needed to
calculate the scattering length of electrons on the corresponding
defect in 3D metal. The obtained analytical expressions are
convenient for analysis of the results of photoionization
experiments. In particular, the concentration of internal vacancies
in a cluster formally can be estimated near the melting point using
these expressions. The jellium and \emph{ab initio} calculations
\cite{Landman-95,Akola-2000,Kanhere,Itoh} reveal a rather subtle
interplay between geometric and electronic shell effects, and
evidences that the quantum mechanical description of the metallic
bonding is crucial for understanding quantitatively the variation in
melting temperatures observed experimentally for free clusters.

There is no doubt that the site of vacancies in the bulk, near or on
the surface \cite{Itoh} will affect the characteristics of the
cluster differently. And without a doubt it is difficult to
determine their contribution separately. However, due to the fact
that the concentration of equilibrium internal vacancies in 3D
metals exponentially depends on the inverse temperature, the surface
energy and work function will contain, in addition to the linear
temperature dependences associated with thermal expansion, also the
exponential dependence which is easier to observe near the melting
of 3D metals. Due to the fact that the energy of the vacancy
formation depends on its site in the cluster (on the surface it is
approximately half the bulk), perhaps we should expect the weak
exponential temperature dependences of their concentrations and,
correspondingly, exponentional temperature-dependent contribution to
both IP and EA. The ability to distinguish and separate these
dependences for large clusters or island films depends of course on
the accuracy of the experiment.

The Fowler-Nordheim theory is applicable for planar field electron
emission. As a modification, we can propose the use of the
temperature and vacancy dependence of the electron work function.
For special electrode geometry, adaptation of the theory is also
possible in an analytical form (see, for example, \cite{Yuasa}).

The self-consistent calculations of the electronic profiles for
perfect clusters and clusters with a vacancy allowed us to determine
the total energy of the neutral and charged defective cluster and
then calculate the dissociation, cohesion, vacancy-formation
energies,  the ionization potential, and electron affinity as well
as the electrical capacitance.

The results of calculations for Na and Al are compared with the
asymptotics and results for defect-free clusters. The ionization
potential for the smallest cluster with a vacancy is greater
 than for a perfect cluster (approximately 0.1 eV for Na and 0.5
eV for Al). A maximum difference is observed at the transition from
a completely filled shell to the empty one. As $N$ increases, this
difference disappears.

Magic numbers of atoms for perfect clusters and clusters with
monovacancy are different, especially for Al. Normalized electrical
capacities of clusters always exceed unity and contain quantum size
fluctuations. In this case, for defective clusters with partially
filled electronic shells capacities are significantly larger than
for perfect clusters.

The size dependence of the cohesion energy contains local maxima.
Clusters corresponding to them are more stable, that is, have
large-scale binding, dissociation, and vacancy-formation energies
than their neighbors. For small clusters, such maxima appear at the
complete filling of the next electron shell. The positions of the
maxima for defect and defect-free clusters are different, which is
due not only to the difference in their sizes, but also due to the
character of the behavior of the electron wave functions.

The energy of cohesion of charged cluster anions and cations is
different from the cohesive energy of neutral clusters. Excessive
positive charge leads to a decrease in energy through increased
forces of electrostatic repulsion, and excess negative charge leads
to the opposite effect.

The quantum-size dependences of the vacancy-formation energies in
the Schottky and the ``bubble blowing'' scenarios, and their
asymptotic tendencies were determined. Strong size fluctuations in
the entire cluster size range were found. Size asymptotics for these
two mechanisms are different from each other and are weakly
dependent on the number of atoms in the cluster. The nature of the
size dependence of vacancy-formation energies from excess charge in
a cluster is determined by the behavior of the ionization potential
of the cluster and the electron affinity.

With the increase in $N$ the dissociation energy is either
increases, or has a local minimum (in the areas between the maxima),
while the vacancy-formation energies decrease monotonically.

Figs. 7 -- 9 indicate that all characteristics of charged and
neutral clusters differ from each other. It is reasonable to assume
that charging might control the melting temperature of the clusters
(see experimental size dependent melting temperature of anions and
cations Al$_{35-70}$ at Fig. 4 in \cite{Jarrold-2009}).

In this model, the relaxation of the cluster volume was not taken
into account. In the limit of large clusters, the effect of
self-compression on the ionization potential is analytically
described in \cite{iakPog-1995,keiPog-1996}, and for small clusters
is numerically investigated in a spherically averaged
pseudopotential model \cite{Vieira}. Relaxation of ionic
distribution in the cluster will lead to a decrease in the total
energy. More coherent, but also more labor-intensive, are \emph{ab
initio} methods with the selection of the coordinates of ions under
the minimum condition of the total energy of the cluster. Such a
procedure is implemented in \cite{Jarrold-2009} only for
Al$_{30-70}$ clusters, i.e., clusters with a small atoms number.

\acknowledgments{We are grateful to Walter V. Pogosov for reading
the manuscript.}

\begin{appendix}

\section{Vacancy quantum  correction}

The straightforward estimation for $\left\langle \delta
V\right\rangle_{R_{N,v}}$ in (\ref{k24}) gives the surprising
result:
\begin{equation}
\left\langle \delta V\right\rangle_{R_{N,v}} \approx \left\langle
\left(R_{v}\frac{ \nabla u_{\rm WS}}{u_{\rm
WS}}\right)\left(R_{N,v}\psi\nabla \psi\right)\right\rangle \approx
\frac{\hbar ^{2}}{mR_{v}^{ 2}} \frac{R_{v}}{R_{N,v}}, \label{k26}
\end{equation}
where $\langle...\rangle$ denotes the integration over the cluster
volume. It seems as if $\left\langle \delta
V\right\rangle_{R_{N,v}}$ is proportional to $(R_{v}R_{N,v})^{-1}$.
In this case the hierarchy of terms in our expansion (\ref{k22})
would be broken, because the previous term is proportional to
$R_{N,v}^{-2}$. However, below we shall demonstrate the emergence of
an extra factor to $\left\langle \delta V\right\rangle_{R_{N,v}}$
which is proportional to $\xi \left(R_{v}/R_{N,v}\right)$,
\begin{equation}
\xi=\frac{L_{v}}{R_{v}}\equiv \frac{L_{v}}{r_{0}}c_{v}^{1/3}\ll 1.
\label{xi}
\end{equation}
It is a result of the integration in Eq. (\ref{k24}) over angles
\cite{PR-2017}.

Due to the fact that the perturbation $\delta V(r)$ occurs on a
scale of supercell, in (\ref{k24}), it will be reasonable to proceed
to integration over supercell and use the Green formula:
\begin{multline}
\left\langle \delta V\right\rangle_{R_{N,v}}=-\frac{\hbar ^{
2}}{2m}\sum\limits_{ i=1}^{ N_{v}}\int \limits_{\Omega_{i}}
d\textbf{r}\:\nabla \Big\{\ln \left[u_{\rm WS}(\varrho)
\right]\Big\} \nabla\psi^{2}(r)
\\
=-\frac{\hbar ^{
2}}{2m}\sum\limits_{i=1}^{N_{v}}\left[\ln u_{\rm WS}
\Big|_{\varrho=R_{v}}\oint\limits_{S_{i}} d{\bf S}\:\nabla\psi^{
2}\right.
\\
- \left.\int\limits_{\Omega_{i}} d\textbf{r}\:\ln u_{\rm WS}\nabla^{
2} \psi^{ 2} \right], \label{k27}
\end{multline}
where $\Omega_{i}$ is a volume of space under condition
$L_{v}<\varrho <R_{v}$.

\begin{figure}[!t!b!p]
\centering
\includegraphics [bb = 27 294 567 794,width=0.48\textwidth] {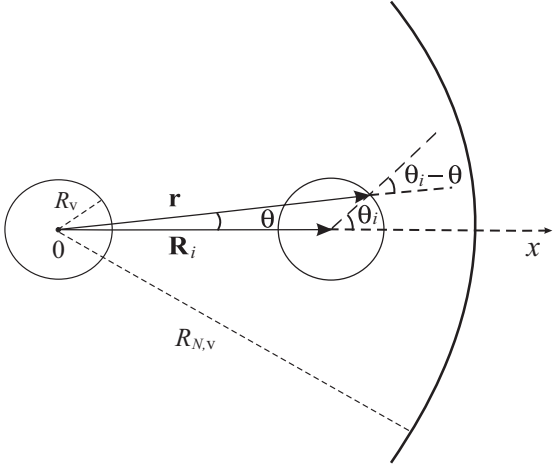}
\caption{The scheme for calculation of the integral over surface of
\emph{i}th cell.} \label{Fig.10}
\end{figure}

Let us introduce the gradient expansion of the squared wave function
$\psi ^{ 2}(r)$ near the point $\textbf{R}_{i}$,
\begin{multline}
\psi^{2}(r)=\psi^{2}(r)\Big|_{r=R_{i}}+\nabla\psi^{2}(r)\Big|_{r=R_{i}}
(\textbf{r}-\textbf{R}_{i})
\\
+ \frac{1}{2}\nabla^{2}\psi^{
2}(r)\Big|_{r=R_{i}}(r^{2}-2\textbf{r}\textbf{R}_{i}+R_{i}^{2})
+\ldots . \label{k28}
\end{multline}
Using expansions of derivatives over the small parameter $\varrho
/R_{ i}\equiv R_{v}/R_{ i}$ we can take the derivatives over
$\textbf{r}$ at the centers of cells $\textbf{R}_{i}$,
\begin{multline}
\nabla\psi ^{ 2}(r)\Big|_{ R_{ i}\gg R_{v}}= \left[\frac{d\psi ^{
2}(R_{ i})}{dR_{ i}}+
\frac{d^{ 2}\psi ^{ 2}(R_{ i})}{dR_{ i}^{ 2}}%
\frac{R_{v}}{R_{ i}} \right]\frac{{\bf r}}{r}
\\
= \frac{d\psi ^{ 2}
(R_{ i})}{dR_{ i}} \left[ 1+O\left( \frac{R_{v}^{ 2}}{R_{ i}^{
2}}\right) \right] \frac{{\bf r}}{r}, \label{k29}
\end{multline}
and
\begin{equation}
\nabla^{ 2} \psi ^{ 2}(r)\Big|_{ R_{ i}\gg R_{v}}=\nabla^{ 2} \psi
^{ 2}(r)\Big|_{ r=R_{ i}}+ O\left(R_{v}^{ 3}/R_{ i}^3\right).
\label{k30}
\end{equation}
Then
\begin{multline}
\left\langle \delta V\right\rangle_{R_{N,v}} =-\frac{\hbar ^{
2}}{2m}\sum\limits_{ i=1}^{ N_{v}}\left[ \frac{d\psi^{2}(R_{
i})}{dR_{ i}}\left(\ln u_{\rm
WS}\right)\Big|_{r=R_{v}}\oint\limits_{S_{i}}d{\bf S}\:\frac{{\bf
r}}{r} \right.
\\
\left.- \nabla^{ 2}\psi^{ 2}(r)\Big|_{r=R_{i}}
\int\limits_{\Omega_{i}} d\textbf{r}\:\ln u_{ \rm WS}\right].
\label{k31}
\end{multline}
Using expression (\ref{k15}), we obtain
\begin{equation}
\left(\ln u_{\rm WS}\right)\Big|_{r=R_{v}}=- \frac{3}{2}( \xi -\xi
^{ 2}) + O(\xi ^{ 3}).
  \label{k32}
\end{equation}
The terms under the sum in Eq. (\ref{k31}) have the opposite signs.

The integral over the surface of \emph{i}th cell is evaluated
exactly (see Fig. 10),
\begin{multline}
\oint\limits_{ S_{ i}}d{\bf S}\:\frac{{\bf r}}{r} = 2\pi R_{v}^{
2}\int \limits_{ 0}^{ \pi} d\theta\: \cos ( \theta_{i}-\theta) \sin
\theta _{i}
\\
=2\pi R_{v}^{ 2}\left\{
\begin{array}{ll}
   1-\frac{1}{3}\frac{R_{ i}}{R_{v}},  \quad & R_{i}<R_{v}, \\
   \frac{2}{3}\frac{R_{v}}{R_{ i}}, \quad & R_{i}>R_{v}. \\
\end{array}
\right. \label{k33}
\end{multline}

Hence, the integration over angles at $R_{i}\gg R_{v}$ gives the
additional power of $R_{v}/R_{i}$. Now, taking account for the
factor $R_{v}/R_{i}$ from Eq. (\ref{k33}), we replace in Eq.
(\ref{k31}) summation by integration and obtain
\begin{equation}
\sum\limits_{ i=1}^{ N}\frac{d\psi^{ 2}}{dR_{ i}} \frac{1}{R_{i}}=
\frac{3}{R_{v}^{3}}\int\limits_{0}^{R_{N}} dr \: r \frac{d\psi^{
2}(r)}{dr} =-\frac{3D_{0}}{2R_{v}^{ 3}R_{N,v}^{2}}, \label{k34}
\end{equation}
$$
D_{0}=-\int\limits_{0}^{ \pi} d\varphi\left( \frac{\sin
2\varphi}{2\varphi}-\frac{\sin ^{ 2}\varphi} {\varphi^{
2}}\right)\approx 0.71.
$$

Next, after the change of the summation over cell numbers to the
integration over the cluster volume it is possible to reduce the
second term in (\ref{k31}) to a surface integral. The integral
vanishes due to the boundary condition (\ref{k21}). Finally,
 using Eqs. (\ref{k33}) and (\ref{k34}), we have first nonzero term
\begin{equation}
\left\langle \delta V\right\rangle_{R_{N,v}}=-\frac{\hbar ^{2}\pi
^{2}}{2mR_{N,v}^{2}} D_{1}\xi +O\left(\frac{\xi^{2}}{R_{N,v}^{2}},
\frac{R_{v}^{ 3}}{R_{N,v}^{3}}\right), \label{k35}
\end{equation}
where $D_{1}=12D_{0}/\pi$. It is worth noting that the excluded
volumes inside the supercells contribute to neglected terms $\sim
\xi^{3}$.

\end{appendix}

\end{document}